\documentclass[12pt]{article}

\usepackage[usenames,dvipsnames,svgnames,table]{xcolor} 
\usepackage{kotex}
\usepackage{graphicx}
\usepackage{epsfig}
\usepackage{amssymb, amsmath, mathtools}
\usepackage{verbatim}
\usepackage{harvard} 
\citationstyle{dcu}
\usepackage[affil-it]{authblk}
\usepackage{mathrsfs}
\usepackage{here}
\usepackage{makecell}
\usepackage{tikz}
\usepackage{enumerate}
\usepackage[shortlabels]{enumitem}
\usepackage{yfonts}
\usepackage{comment}
\usepackage{longtable}
\usepackage{multirow} 
\usepackage{algorithm,algpseudocode}
\usepackage{subcaption}  
\usepackage{arydshln}

\usepackage{hyperref}
\usepackage[colorinlistoftodos, textsize=scriptsize]{todonotes} 

\usepackage[framemethod=TikZ]{mdframed}
\mdfdefinestyle{JL}{%
	linecolor=blue,
	outerlinewidth=1pt,
	roundcorner=2pt,
	innertopmargin=0.1\baselineskip,
	innerbottommargin=0.1\baselineskip,
	innerrightmargin=2pt,
	innerleftmargin=2pt,
	backgroundcolor=orange!100!white}
\newenvironment{JL}{\begin{mdframed}[style=JL]  \footnotesize} { \end{mdframed}}   
\newcommand{\bJ}{\begin{JL}}
	\newcommand{\eJ}{\end{JL}}

\includecomment{note} 

\newtheorem{theorem}{Theorem}[section]
\newtheorem{lemma}[theorem]{Lemma}
\newtheorem{definition}[theorem]{Definition}
\newtheorem{proposition}[theorem]{Proposition}
\newtheorem{corollary}[theorem]{Corollary}

\newtheorem{remark}[theorem]{Remark}

\everymath{\displaystyle}

\newcommand{\half}{\frac{1}{2}}

\newcommand{\ra}{\rightarrow}

\newcommand{\be}{\begin{equation}}
\newcommand{\ee}{\end{equation}}
\newcommand{\bea}{\begin{eqnarray*}}
	\newcommand{\eea}{\end{eqnarray*}}
\newcommand{\bean}{\begin{eqnarray}}
\newcommand{\eean}{\end{eqnarray}}
\newcommand{\ben}{\begin{enumerate}}
	\newcommand{\een}{\end{enumerate}}
\newcommand{\bi}{\begin{itemize}}
	\newcommand{\ei}{\end{itemize}}
\newcommand{\brem}{\begin{remark}}
	\newcommand{\erem}{\end{remark}}
\newcommand{\bcen}{\begin{center}}
	\newcommand{\ecen}{\end{center}}
\newcommand{\bsv}{\begin{semiverbatim}}
	\newcommand{\esv}{\end{semiverbatim}}

\newcommand{\bt}{\begin{theorem}}
	\newcommand{\et}{\end{theorem}}
\newcommand{\bl}{\begin{lemma}}
	\newcommand{\el}{\end{lemma}}
\newcommand{\bd}{\begin{definition}}
	\newcommand{\ed}{\end{definition}}
\newcommand{\bc}{\begin{corollary}}
	\newcommand{\ec}{\end{corollary}}
\newcommand{\bp}{\begin{proposition}}
	\newcommand{\ep}{\end{proposition}}

\newcommand{\bff}{ \mathbf{f}}

\newcommand{\bfu}{ \mathbf{u}}

\newcommand{\bfx}{ \mathbf{x}}
\newcommand{\bfy}{ \mathbf{y}}

\newcommand{\bfA}{ \mathbf{A}}
\newcommand{\bfB}{ \mathbf{B}}
\newcommand{\bfC}{ \mathbf{C}}
\newcommand{\bfD}{ \mathbf{D}}

\newcommand{\bfO}{ \mathbf{O}}
\newcommand{\bfX}{ \mathbf{X}}
\newcommand{\bfY}{ \mathbf{Y}}  
\newcommand{\bfZ}{\mathbf{Z}}

\newcommand{\bbR}{ \mathbb{R}}

\newcommand{\btheta}{ \boldsymbol{\theta}}

\newcommand{\bed}{\begin{itemize}}
	\newcommand{\eed}{\end{itemize}}

\newcommand{\bfg}{ \mathbf{g}}
\newcommand{\bfv}{ \mathbf{v}}
\newcommand{\bfI}{ \mathbf{I}}
\newcommand{\bfJ}{ \mathbf{J}}
\newcommand{\bfH}{ \mathbf{H}}
\newcommand{\bvarepsilon}{ \boldsymbol{\varepsilon}}
\newcommand{\bfeta}{ \boldsymbol{\eta}}

\newcommand{\bzero}{\boldsymbol{0}}

\graphicspath{{fig/}} 

\title{Laplace-aided Variational Inference for Differential Equation Models}
\author{Hyunjoo Yang and Jaeyong Lee}
\affil{Department of Statistics \\ Seoul National University}

\begin{document}

\maketitle

\begin{abstract}
An ordinary differential equation (ODE) model, whose regression curves are a set of solution curves for some ODEs, poses a challenge in parameter estimation.  The challenge, due to the frequent absence of analytic solutions and the complicated likelihood surface, tends to be more severe especially for larger models with many parameters and variables. \citeasnoun{yang2021variational} proposed a \textit{state-space model with variational Bayes (SSVB)} for ODE, capable of fast and stable estimation in somewhat large ODE models. The method has shown excellent performance in parameter estimation but has a weakness of underestimation of the posterior covariance, which originates from the mean-field variational method. This paper proposes a way to overcome the weakness by using the Laplace approximation. In numerical experiments, the covariance modified by the Laplace approximation showed a high degree of improvement when checked against the covariances obtained by a standard Markov chain Monte Carlo method. With the improved covariance estimation, the SSVB  renders fairly accurate posterior approximations.

\end{abstract}

\section{Introduction}
\label{s:Lap_intro}
An ordinary differential equation (ODE) model assumes that the regression curves fitting the observed data are a set of solution curves for some specific ODEs, 
\begin{align*}
\dot{\bfx}(t)=\bff(\bfx(t),t\ ;\btheta),\qquad t\in[0,T],
\end{align*} 
with the ODE parameters $\btheta \in \Theta \subset \bbR^q$. 
Despite the good interpretability of ODE itself, parameter estimation from the observed data poses a challenge due to the frequent absence of analytic solutions and the complicated likelihood surface. In particular, as the model contains more parameters and variables, these tendencies become more severe, making fast and accurate parameter estimation increasingly difficult.

For fast inference, ODE parameter estimating algorithms have been developed in a way that avoids the computation of numerical solutions. The two-step approach, which the frequentist methods are mainly based on, approximate the ODE solutions by other non-linear regression such as spline expansion \cite**{ramsay2005functional,ramsay2007parameter}, local polynomial regression \cite**{liang2008parameter,liang2010estimation}, and so on. In the Bayesian framework, similarly, the methods using Gaussian processes as a regression curve have been developed \cite**{calderhead2008accelerating,dondelinger2013ODE,wang2014gaussian}, while there are other methods to improve Markov chain Monte Carlo (MCMC) using parallel tempering algorithm \cite{campbell2012smooth}, sequential Monte Carlo \cite**{lee2018inference}. An important limitation of the existing algorithms is that the performance evaluations and comparisons have been made virtually under small models, such as the FitzHugh-Nagumo model.

\citeasnoun{yang2021variational} proposed a Bayesian method, a \textit{state-space model with variational Bayes (SSVB)}, capable of fast and stable estimation in somewhat large ODE models. The two main strategies are an approximation to a state-space model and the variational Bayes. First, the SSVB relaxes the original ODE model of $p$ variables to the following state-space model with the tuning parameter $\tau$:
\begin{align}
\label{SSM}
\begin{split}
\bfy_i&=\bfx_i+\bvarepsilon_i,\quad \bvarepsilon_i\overset{iid}{\sim}\text{N}(\bzero,\lambda^{-1}\bfI_p), \quad i=0,1,\dots,n,\\
\bfx_{i+1}=&\ \bfg(\bfx_i,t_i,\btheta)+\bfeta_i,\quad \bfeta_i\overset{iid}{\sim}\text{N}(\bzero,\tau\bfI_p), \quad i=0,1,\dots,n-1.
\end{split}
\end{align}
In the model, the 4th-order Runge-Kutta method was chosen as an approximating function $ \bfg(\cdot) $ as follows: 
\begin{align}
\label{Runge}
\begin{split}
\bfg(\bfx_i^*,t_i,\btheta)=&\ \bfx_i^*+\frac{1}{6}(K_{i1}+2K_{i2}+2K_{i3}+K_{i4}),\\
K_{i1}=&\ h_{i+1}\cdot \bff(\bfx_i^*,t_i;\btheta),\\
K_{i2}=&\ h_{i+1}\cdot \bff(\bfx_i^*+\frac{1}{2}K_{i1},t_i+\frac{1}{2}h_{i+1};\btheta),\\
K_{i3}=&\ h_{i+1}\cdot \bff(\bfx_i^*+\frac{1}{2}K_{i2},t_i+\frac{1}{2}h_{i+1};\btheta),\\
K_{i4}=&\ h_{i+1}\cdot \bff(\bfx_i^*+K_{i3},t_i+h_{i+1};\btheta),
\end{split}
\end{align}
where $h_{i+1} = t_{i+1} - t_i$ for $i = 0,1, \ldots, n-1$. The relaxed model allows fast estimation by avoiding computations of a complete numerical solution in the likelihood. Second, as a Bayesian method, the SSVB algorithm exploits the variational Bayes method for computing the posterior. The variational Bayes method also enables fast estimation by converting the inference of the posterior into an optimization problem rather than an MCMC-like sampling method. Furthermore, it increases the accuracy of the inference since the concentrativeness of mean-field approximation makes it advantageous to estimate good combinations of the initial values $\bfx_0$ and the ODE parameters $\btheta$, which is crucial to the reproduction of the true ODE curves. For more details, see \citeasnoun{yang2021variational}.

Indeed, their simulation studies showed the SSVB's fast and accurate performance with strong stability even in a large model with more than 30 parameters. Especially, it was markedly superior in reproducing the ODE curves while all the other competing estimators \cite**{haario2006dram,ramsay2007parameter,hoffman2014no,lee2018inference}  have not provided valid inferences for the same large model. Taking advantage of it, the algorithm was also successfully applied to the time-varying SIR model \cite{yang2021variational} with many parameters for the COVID-19 epidemic data.

Despite the excellent performance, however, the SSVB based on the mean-field variational method still has a well-known weakness: the underestimation of posterior covariances. The variational distribution in which all the parameters are mutually independent provides no information about the correlations between the different parameters, and above all, from the structure of the objective function, it underestimates the variances representing the uncertainty of the inference \cite**{blei2017variational}. The good performance of the SSVB could be seen as a trade-off for this over-approximations.

To improve the underestimated variance problem caused by the mean-field assumption, \citeasnoun**{giordano2015linear} proposed a method, \textit{linear response variational Bayes} (LRVB), by generalizing linear response methods originated in statistical physics. However, the LRVB is difficult to apply to the SSVB because the state-space model has a strong dependency between neighboring latent variables. Since the method is affected by the second moment estimates from the variational distribution, the more severe correlations between variables exist, the worse its performance gets.

This paper proposes a solution using the Laplace approximation, a kind of density approximating method, for the problem of variance underestimation. Using the second derivatives at the mode, the density function is approximated to that of a multivariate normal distribution. The idea of applying it to the SSVB is to treat the SSVB's parameter estimates as the mode of the posterior density and to suggest the Laplace approximated one as a better covariance estimate.  In experiments, the modified covariance showed a high degree of improvement when checked against the covariances obtained by standard MCMC methods. With the improved covariance estimation, the SSVB can be a promising ODE estimation algorithm.

Section \ref{s:Lap_method} describes the Laplace approximation method applied to the SSVB. In addition to the simple application, the Laplace approximation applied to the original ODE model rather than the relaxed model (\ref{SSM}) is also presented as a secondary method. The experimental results comparing the covariance estimates are provided in Section \ref{s:Lap_simul}. In Section \ref{s:Lap_appl}, the application results to the time-varying SIR model in \citeasnoun{yang2021variational} for the COVID-19 data are provided as a real-world data analysis. Discussions are given in Section \ref{s:Lap_disc}, whereas details of computations are relegated to the Appendix.

\section{Laplace approximation for the posterior covariance} 
\label{s:Lap_method}
In terms of the probability density function, the Laplace approximation is an approximation to Gaussian distribution. It uses the second derivative at the mode of the density as the precision matrix, the inverse of the covariance matrix. Following  \citeasnoun{mackay2003information}, let $ p^*(\cdot)$ be an unnormalized density of $k$-dimensional random vector $\bfX$, which has the mode $\bfx^*\in\mathbb{R}^k$. Taylor's expansion for $\log p^*(\bfx)$ about the point $\bfx^*$ produces the following approximate equation:
$$ \log p^*(\bfx)\approx \log p^*(\bfx^*)-\half(\bfx-\bfx^*)^T \bfH (\bfx-\bfx^*) + \cdots, $$
where $\bfH$ is a $k\times k$ symmetric matrix whose elements are
$$ \bfH_{ij}=-\frac{\partial^2}{\partial x_i \partial x_j}\log p^*(\bfx)\Bigg|_{\bfx=\bfx^*} .$$
As a result, the unnormalized pdf $ p^*(\bfx) $ is approximated to the multivariate normal distribution
$$ q(\bfx) \propto e^{-\half(\bfx-\bfx^*)^T \bfH (\bfx-\bfx^*)},$$
with the mean vector $ \bfx^* $ and the covariance matrix $\bfH^{-1}$.

The Laplace approximation presupposes that the mode $ \bfx^* $ is known. In applying to the correction of posterior covariance for the SSVB, we use the point estimate from the SSVB, or the variational mean parameter, as the mode. This selection is based on the assumption that the point estimate of the SSVB, which shows high performance in reproducing the solution curve, would be close to the mode with the highest posterior density. 

Since the SSVB algorithm uses a state-space model relaxed from the ODE model, two posterior distributions can be considered for the Laplace approximation.  One is from the SSVB, the state-space model, and the other is from the original ODE model.

\subsection{Laplace approximation of the relaxed model}\label{subsec:lap_to_relaxed}
When the prior is given by
\begin{align}
\label{prior}
\begin{split}
\lambda\sim& \ \text{Gamma}(A_0,B_0), \\
\theta_k\sim& \ \text{Unif}(a_{0k},b_{0k}), \ \text{for }k=1,\dots,q, \\
\bfx_{0j}\sim& \ \text{Unif}(c_{0j},d_{0j}),\ \text{for }j=1,\dots,p,
\end{split}
\end{align}
the posterior of the SSVB model (\ref{SSM}) can be obtained as follows:
$$
p(\lambda,\btheta,\bfX|\bfY)\propto \lambda^{\frac{p(n+1)}{2}+A_0-1}e^{-B_0\lambda}e^{-\frac{1}{2\tau}\sum_{i=1}^{n} \lVert\bfx_i-\bfg(\bfx_{i-1},\btheta)\rVert^2-\frac{\lambda}{2}\sum_{i=0}^{n}\lVert\bfy_i-\bfx_i\rVert^2}.
$$
For the Laplace approximation, from the log-posterior
\begin{align*}
	L:= -\log p(\lambda,\btheta,\bfX|\bfY)=&-\left( \frac{p(n+1)}{2}+A_0-1 \right)\log\lambda + B_0\lambda \\
	&+ \frac{1}{2\tau}\sum_{i=1}^{n} \lVert\bfx_i-\bfg(\bfx_{i-1},\btheta)\rVert^2+\frac{\lambda}{2}\sum_{i=0}^{n}\lVert\bfy_i-\bfx_i\rVert^2 + C, 
\end{align*}
the second derivative, the Hessian matrix $\bfH$, can be calculated by the chain rule. Some are as follows:
\begin{align*}
&\frac{\partial^2L}{\partial\lambda^2} = \frac{1}{\lambda^2}\left( \frac{p(n+1)}{2}+A_0-1 \right), \\
&\frac{\partial^2L}{\partial\btheta\partial\lambda} = \bzero, \\
&\frac{\partial^2L}{\partial\btheta\partial\btheta^T} = -\frac{1}{\tau} \sum_{i=1}^{n} \left[\sum_{j=1}^p (x_{ij}-g_j(\bfx_{i-1},\btheta))\cdot \frac{\partial^2 g_j(\bfx_{i-1},\btheta)}{\partial\btheta\partial\btheta^T} - \bfJ_{\bfg\text{ wrt }\btheta}(\bfx_{i-1},\btheta)^T\bfJ_{\bfg\text{ wrt }\btheta}(\bfx_{i-1},\btheta) \right], \\
&\frac{\partial^2L}{\partial\bfx_0\partial\btheta^T} = -\frac{1}{\tau} \left[ \sum_{j=1}^p (x_{1j}-g_j(\bfx_{0},\btheta))\cdot \frac{\partial^2 g_j(\bfx_{0},\btheta)}{\partial\bfx\partial\btheta^T} - \bfJ_{\bfg\text{ wrt }\bfx}(\bfx_{0},\btheta)^T\bfJ_{\bfg\text{ wrt }\btheta}(\bfx_{0},\btheta) \right] .
\end{align*}
Here, $\bfJ_{\bfg\text{ wrt }\bfx}(\cdot)$ and $\bfJ_{\bfg\text{ wrt }\btheta}(\cdot)$ stand for the Jacobian matrices of $\bfg(\cdot)$ with respect to $\bfx$ and $\btheta$, respectively (for the detailed computation, see Appendix of \citeasnoun{yang2021variational}). For the simplicity of notation, the argument $t_i$ in $\bfg(\bfx_i,t_i,\btheta)$ corresponding to the time of $\bfx_i$ is omitted. The full equations of the second derivatives are given in Appendix \ref{2ndFull}. The calculational details for the Hessian matrix of $\bfg$ with respect to $(\bfx^T,\btheta^T)^T$ in the equations are given in Appendix \ref{Hessian}. As mentioned above, the $\bfH^{-1}$ at the estimates $(\hat{\btheta},\hat{\bfX} )$ from the SSVB is regarded as the modified covariance.

\subsection{Laplace approximation of the original ODE model}
Naturally, instead of the relaxed model, we can also consider the posterior from the original ODE model 
\begin{align}
\label{ODE model}
\begin{split}
\bfy_i=&\ \bfx_i+\bvarepsilon_i,\quad\bvarepsilon_i\overset{iid}{\sim}\text{N}(\bzero,\lambda^{-1}\bfI_p), \\
\dot{\bfx}(t)&=\bff(\bfx(t),t\ ;\btheta),
\end{split}
\end{align} 
as the likelihood.
Here, $ \bfx_i:=\bfx( t_i;\btheta, \bfx_0) $ for $\ i=0,1,\dots,n $ are the points on the ODE solution curve $\bfx(t;\btheta, \bfx_0) $ determined by the initial values $\bfx_0$ as well as $\btheta$.

With the same prior of (\ref{prior}), the posterior of the original ODE model and the logarithm are:
$$
p(\lambda,\btheta,\bfx_0|\bfY)\propto \lambda^{\frac{p(n+1)}{2}+A_0-1}e^{-B_0\lambda}e^{-\frac{\lambda}{2}\lVert\bfy_0-\bfx_0\rVert^2-\frac{\lambda}{2}\sum_{i=1}^{n}\lVert\bfy_i-\bfx(t_i;\btheta, \bfx_0)\rVert^2}, 
$$
\begin{align*}
	 L:=-\log p(\lambda,\btheta,\bfx_0|\bfY)=&-\left( \frac{p(n+1)}{2}+A_0-1 \right)\log\lambda + B_0\lambda\\
	 &+\frac{\lambda}{2}\lVert\bfy_0-\bfx_0\rVert^2 +\frac{\lambda}{2}\sum_{i=1}^{n}\lVert\bfy_i-\bfx(t_i;\btheta, \bfx_0)\rVert^2 + C.
\end{align*}
The second derivatives of the logarithm for the Laplace approximation can be calculated as follows:
\begin{align*}
&\frac{\partial^2L}{\partial\lambda^2} = \frac{1}{\lambda^2}\left( \frac{p(n+1)}{2}+A_0-1 \right), \\
&\frac{\partial^2L}{\partial\btheta\partial\lambda} = - \sum_{i=1}^{n} [\bfJ_{\bfx_i\text{ wrt }\btheta}]^T (\bfy_i-\bfx(t_i;\btheta, \bfx_0)), \\
&\frac{\partial^2L}{\partial\bfx_0\partial\lambda} = (\bfx_0-\bfy_0) - \sum_{i=1}^{n} [\bfJ_{\bfx_i\text{ wrt }\bfx_0}]^T (\bfy_i-\bfx(t_i;\btheta, \bfx_0)), \\
&\frac{\partial^2L}{\partial\btheta\partial\btheta^T} = -\lambda\sum_{i=1}^{n} \left[\sum_{j=1}^p (y_{ij}-x_{j}(t_i;\btheta, \bfx_0))\cdot \frac{\partial^2 x_{j}(t_i;\btheta, \bfx_0)}{\partial\btheta\partial\btheta^T} - [\bfJ_{\bfx_i\text{ wrt }\btheta}]^T\bfJ_{\bfx_i\text{ wrt }\btheta} \right], \\
&\frac{\partial^2L}{\partial\bfx_0\partial\btheta^T} = -\lambda \sum_{i=1}^{n}\left[ \sum_{j=1}^p (y_{ij}-x_{j}(t_i;\btheta, \bfx_0))\cdot \frac{\partial^2 x_{j}(t_i;\btheta, \bfx_0)}{\partial\bfx_0\partial\btheta^T} - [\bfJ_{\bfx_i\text{ wrt }\bfx_0}]^T\bfJ_{\bfx_i\text{ wrt }\btheta} \right], \\
&\frac{\partial^2L}{\partial\bfx_0\partial\bfx_0^T} =\lambda\bfI_p -\lambda \sum_{i=1}^{n}\left[ \sum_{j=1}^p (y_{ij}-x_{j}(t_i;\btheta, \bfx_0))\cdot \frac{\partial^2 x_{j}(t_i;\btheta, \bfx_0)}{\partial\bfx_0\partial\bfx_0^T} - [\bfJ_{\bfx_i\text{ wrt }\bfx_0}]^T\bfJ_{\bfx_i\text{ wrt }\bfx_0} \right] .
\end{align*}
Here, $ \bfJ_{\bfx_i\text{ wrt }\bfx_0} $ and $ \bfJ_{\bfx_i\text{ wrt }\btheta} $ stand for the Jacobian matrices of $\bfx_i$ with respect to $\bfx_0$ and $\btheta$, respectively. The above computations include the second derivatives of an ODE solution $\bfx(t;\btheta, \bfx_0)$ with respect to $(\btheta^T,\bfx_0^T)^T$. These calculations, also known as sensitivity analysis, can be computed as a solution of an ODE system that is extended from the ODE $\dot{\bfx}(t)=\bff(\bfx(t),t\ ;\btheta)$ in the model. Using results from \citeasnoun{dickinson1976sensitivity} and \citeasnoun{barrio2006sensitivity}, the details of calculation are given in Appendix \ref{sensitivity}. As in the previous case, the posterior covariance estimate is calculated by regarding the point estimate of the SSVB as the mode.

\section{Experimental results}
\label{s:Lap_simul}
To determine whether the proposed covariances are close to the true posterior covariance, we need information about the true one. Since the exact covariance is virtually impossible to calculate due to the ODE model's complex likelihood, sampling methods based on MCMC are used for comparison. 

The two sampling methods used for comparison are the DRAM (delayed rejection \& adaptive metropolis) algorithm of \citeasnoun{haario2006dram} and an HMC (Hamiltonian Monte Carlo) algorithm of \citeasnoun{neal2011mcmc}. The DRAM algorithm is a combination of the delayed rejection (DR) algorithm, which delays the rejection and considers more candidates at each sampling iteration, and the adaptive Metropolis (AM) algorithm, which adapts the covariance of the proposal distribution reflecting the samples so far. The computation is conducted through the \textbf{R} package \textbf{FME}  \cite{soetaert2010inverse}, and a maximum of 2 candidates are considered in each update iteration for DR. The HMC is an MCMC method based on the Hamiltonian dynamics, taking the density function as a potential energy function. The algorithm can be implemented through the \textbf{rstan} package \cite{rstan2020} with the default option for the no-U-turn sampler (NUTS) of \citeasnoun{hoffman2014no}, an adaptive variant of HMC.

Since the purpose is to compare the covariance estimates, the starting values for the above two algorithms are given `nicely'. In large ODE models, MCMC-based methods frequently fail to estimate the parameters, meaning that they even do not reach the main region with a dominant probability of the posterior. This also motivated the development of the SSVB. Therefore, to prevent these fails, the starting values are given as the resulting point estimates from the SSVB algorithm with good performance.

Experiments were conducted on two ODE models, the FitzHugh-Nagumo model and the Lorenze-96 model. For each experiment, correlation structures and variances from the methods were compared.

\subsection{FitzHugh-Nagumo model}
The FitzHugh-Nagumo model 
\begin{align*}
\label{FN}
\begin{split}
\dot{x}_1(t)=&\ \ \theta_3\left(x_1(t)-\frac{1}{3}x^3_1(t)+x_2(t)\right),\\
\dot{x}_2(t)=&-\frac{1}{\theta_3}\bigg(x_1(t)-\theta_1+\theta_2x_2(t)\bigg),
\end{split}
\end{align*}  
with two variables and three parameters, is a popular model in ODE parameter estimation studies. A data set was generated from the true model of $\btheta=(0.2,\ 0.2,\ 3)^T$ and $\bfx_0=(-1,\ -1)^T$ along the 201 equidistant time points $t_0=0,\ t_1=0.1,\ \cdots,\ t_{200}=20$, with the error variance $1/\lambda=0.25$. 

All the methods (algorithms) were run under the same prior in the form of (\ref{prior}). For the DRAM algorithm, a chain of length 1,000, which is every 30th iteration retained by thinning from 30,000 iterations after 5,000 burn-in, was obtained. The NUTS algorithm of HMC, with better movements, also provided a chain of length 1,000, which is every 10th iteration retained by thinning from 10,000 iterations after 5,000 burn-in. For the SSVB, the tuning parameter $\tau$ was set to $0.1^5$. 

\begin{figure}[b!]
	\centering
	\includegraphics[width=0.8\linewidth]{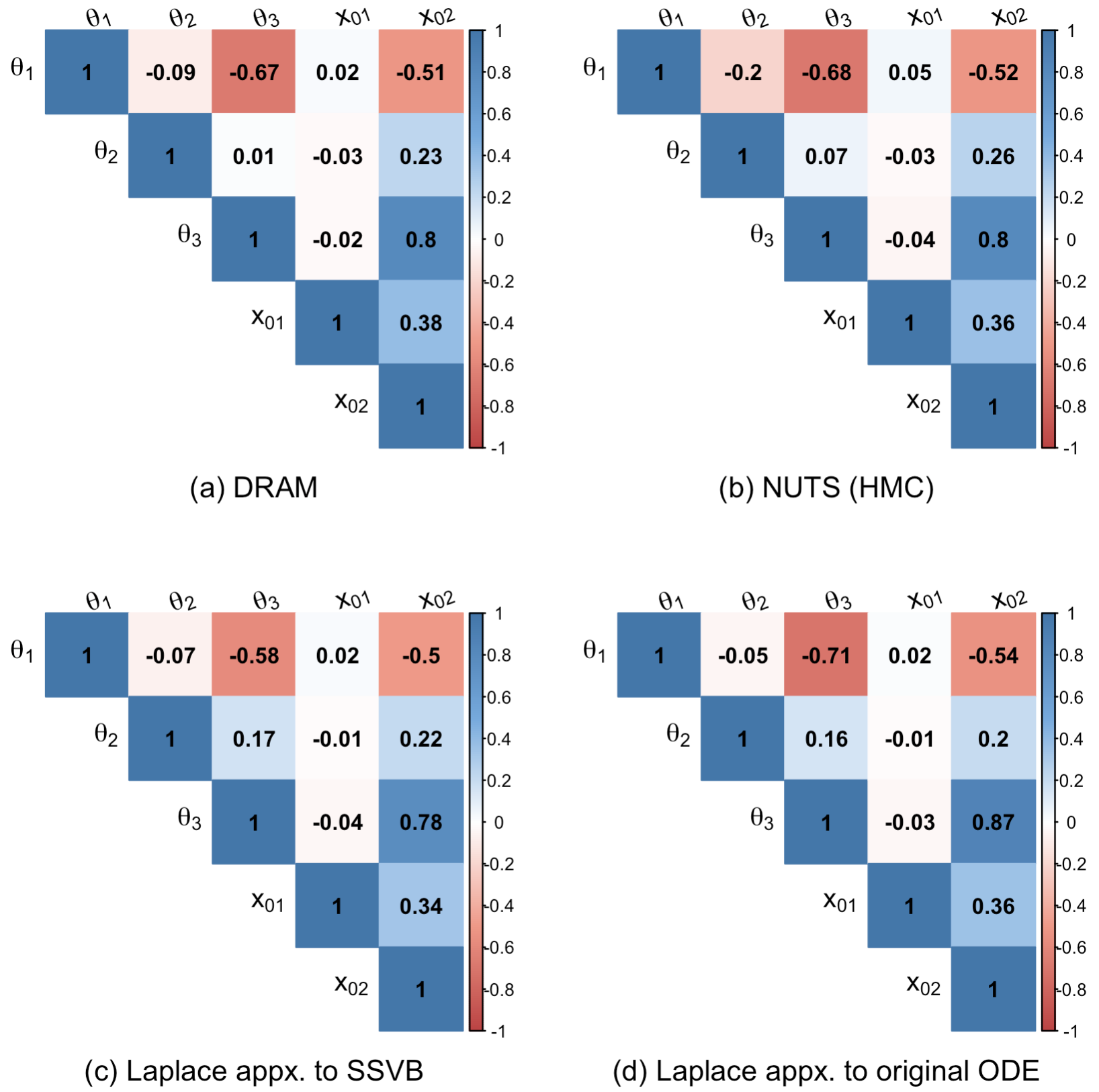}
	\caption{Posterior correlation structures of the FitzHugh-Nagumo model from (a) DRAM, (b) NUTS (HMC), (c) SSVB with Laplace approximation, (d) Original ODE model with Laplace approximation.}
	\label{fig:NagumoCorr}
\end{figure}

Figure~\ref{fig:NagumoCorr} shows the resulting correlation structures. As mentioned earlier, basically, the SSVB method under the mean-field assumption does not provide any information about the correlation between parameters. The correlation matrix would be filled with zeros except for the diagonal. With the Laplace approximation, however, the modified correlation showed a similar pattern to those from the DRAM and NUTS, which were assumed as a standard. The Laplace approximation applied to the original ODE model, which has the same posterior of the DRAM and NUTS, also provided a similar structure. Though the exact correlation structure is unknown, the similarity of the results indicates that the proposed methods render reasonable posterior covariances.

\begin{figure}[t!]
	\centering
	\includegraphics[width=0.7\linewidth]{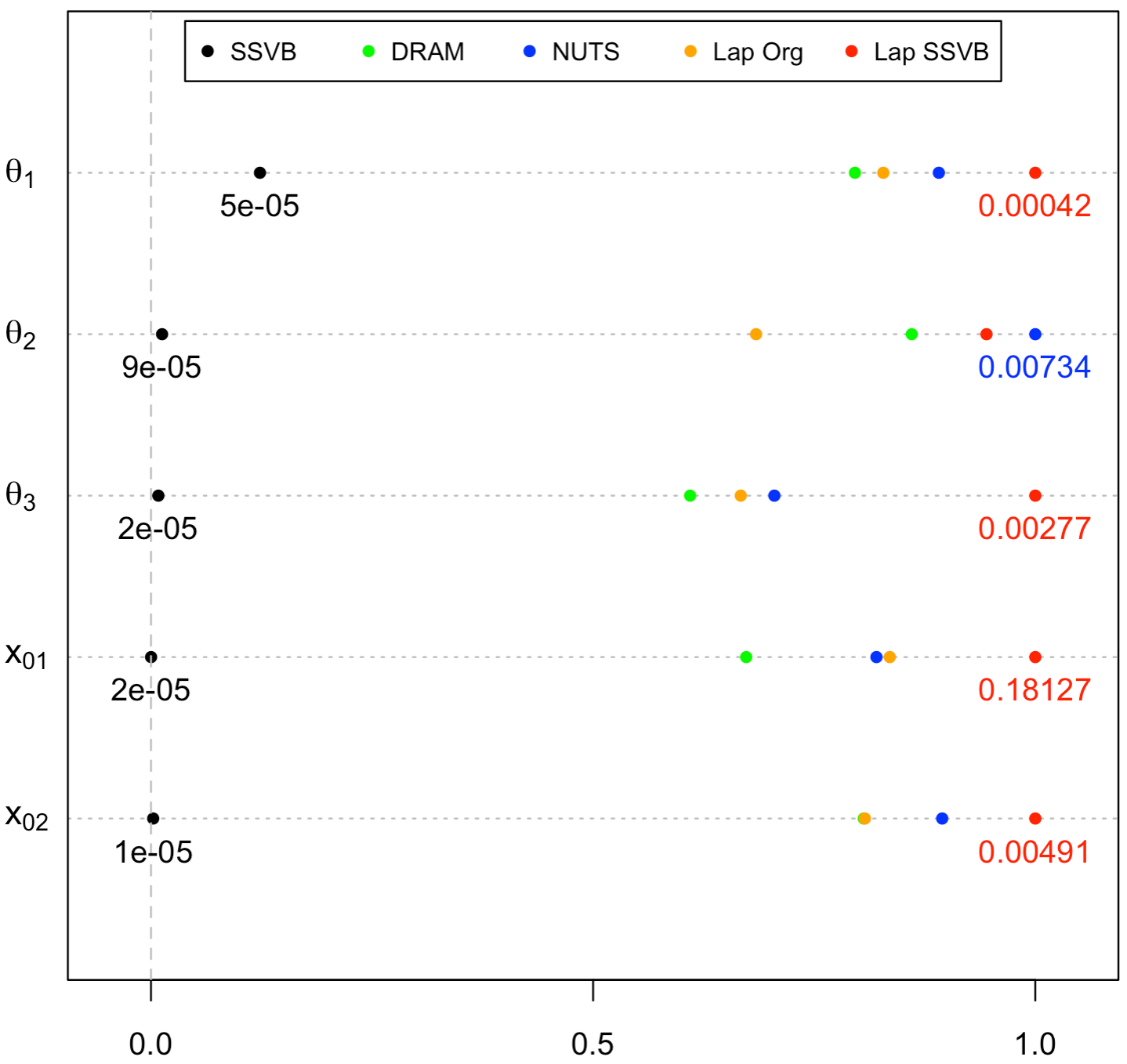}
	\caption{Posterior variance estimates of the FitzHugh-Nagumo model. For intuitive visualization of the scale, the estimates relative to the largest one are plotted on a horizontal line for each parameter.}
	\label{fig:NagumoVar}
\end{figure}

Secondly, the scale of posterior variance, representing the uncertainty of inference, was checked. In order to intuitively visualize how much the underestimated uncertainty of the SSVB was recovered, in Figure~\ref{fig:NagumoVar}, the estimates from the five methods relative to the largest one are plotted on a horizontal line for each parameter. The specific figures for the smallest and largest ones are also added. As expected, the estimates from the SSVB were overwhelmingly smaller than those of the MCMC-based methods, but jumped to a similar scale after being modified by the Laplace approximation. The results of the Laplace approximation on the original ODE model, shown in orange, were also comparable to other methods in scale. 

Looking closely, the SSVB modified by the Laplace approximation generally showed a larger value than the others. This seems to be because the SSVB is based on a relaxed model rather than the original ODE model. The reasoning also supported by the result that the Laplace approximation of the original ODE model provided the closer estimates to the MCMC methods with the same likelihood. What is clear is that the Laplace approximated covariances are much better estimators than the underestimated variance of just the SSVB. In both the correlation structure and the scale of variance, the Laplace approximation performs reasonably well.

\begin{figure}[p]
	\centering
	\begin{subfigure}{0.24\linewidth}
		\includegraphics[width=\linewidth]{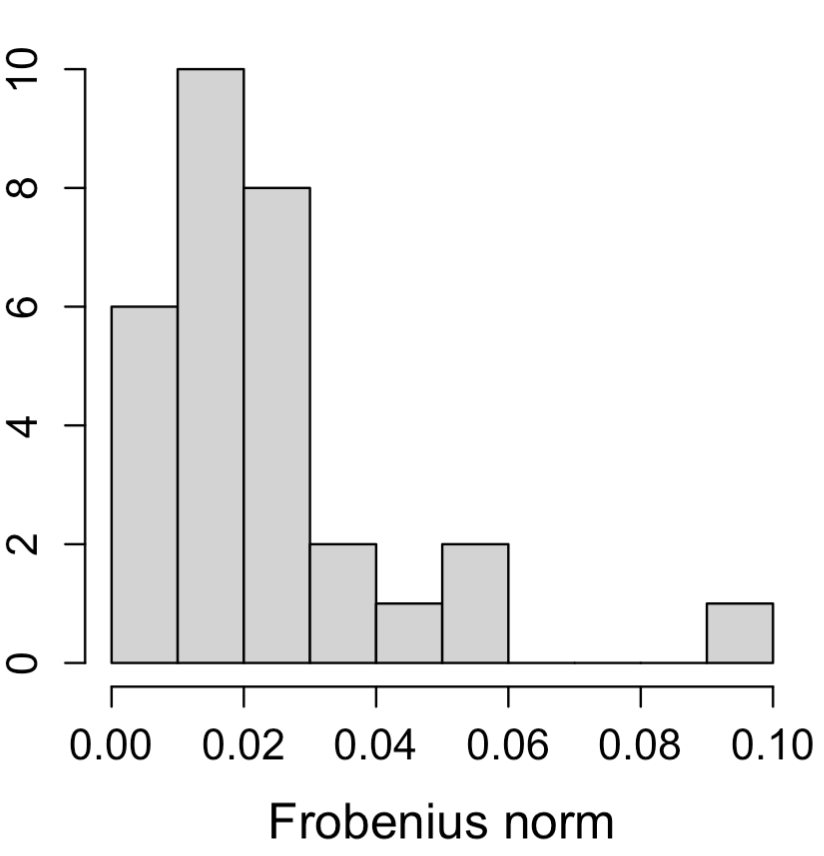} 
		\subcaption{Lap SSVB - DRAM}
	\end{subfigure}
	\begin{subfigure}{0.215\linewidth}
		\includegraphics[width=\linewidth]{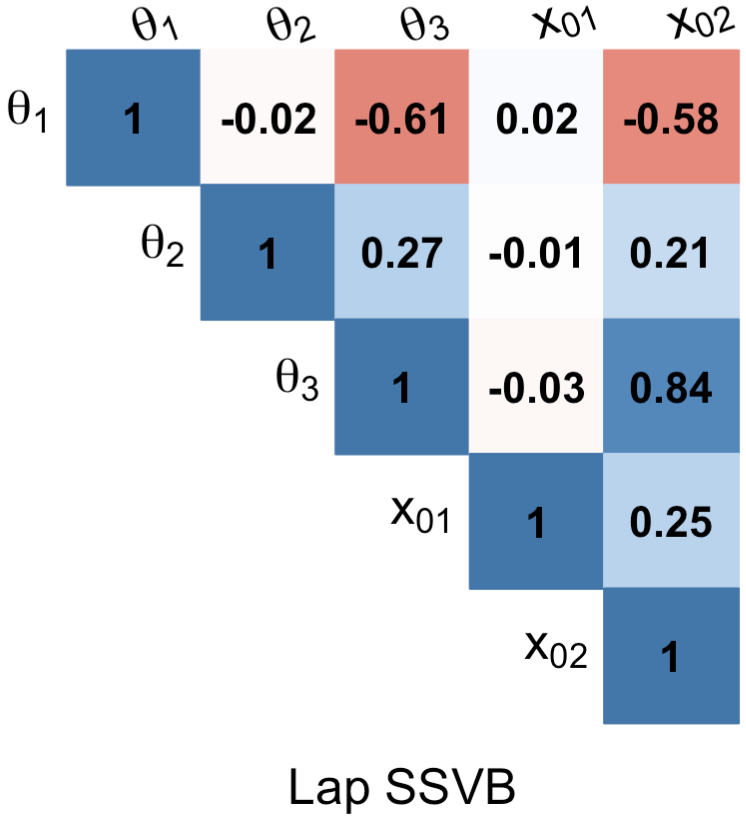}
		\subcaption*{}
	\end{subfigure}
	\begin{subfigure}{0.24\linewidth}
		\includegraphics[width=\linewidth]{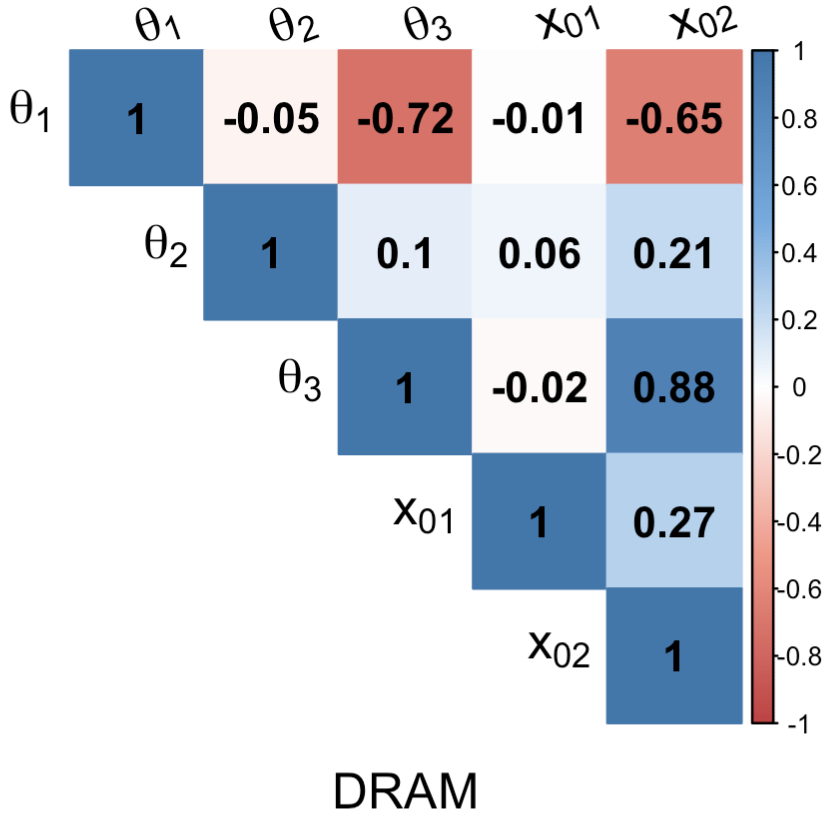}
		\subcaption*{}
	\end{subfigure}
	\ 
	\begin{subfigure}{0.24\linewidth}
		\includegraphics[width=\linewidth]{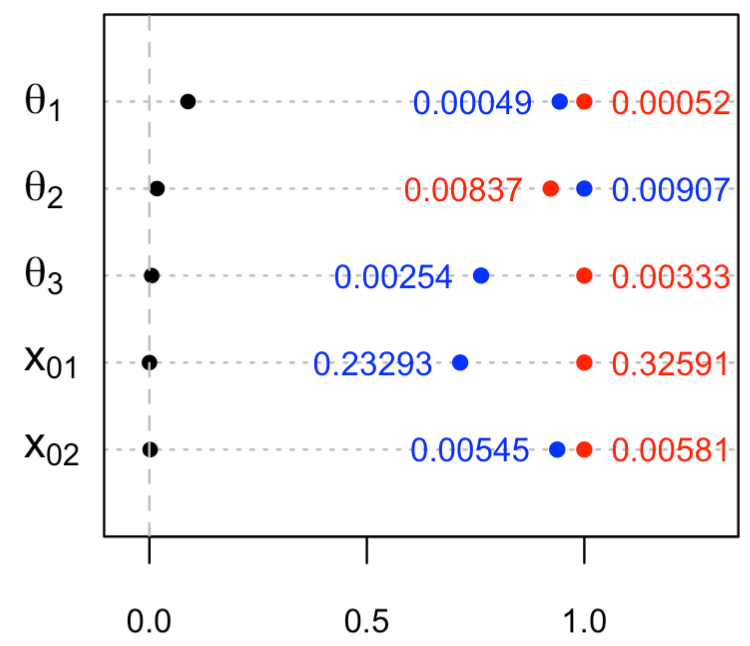}
		\subcaption*{}
	\end{subfigure}\\
	\hrulefill 
	\newline
	\begin{subfigure}{0.24\linewidth}
		\includegraphics[width=\linewidth]{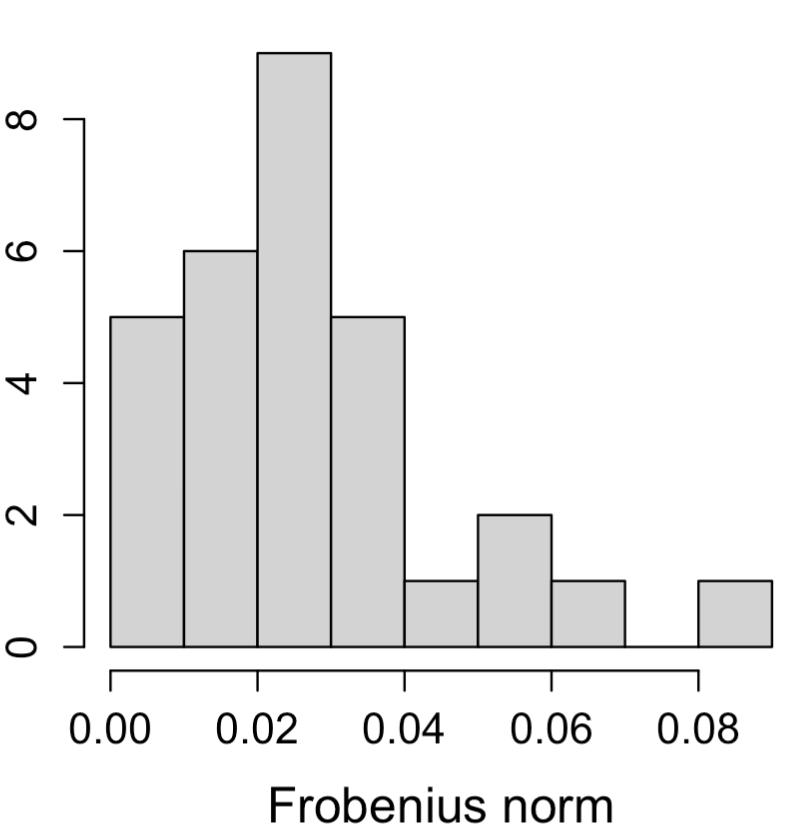}
		\subcaption{Lap SSVB - NUTS}
	\end{subfigure}
	\begin{subfigure}{0.215\linewidth}
		\includegraphics[width=\linewidth]{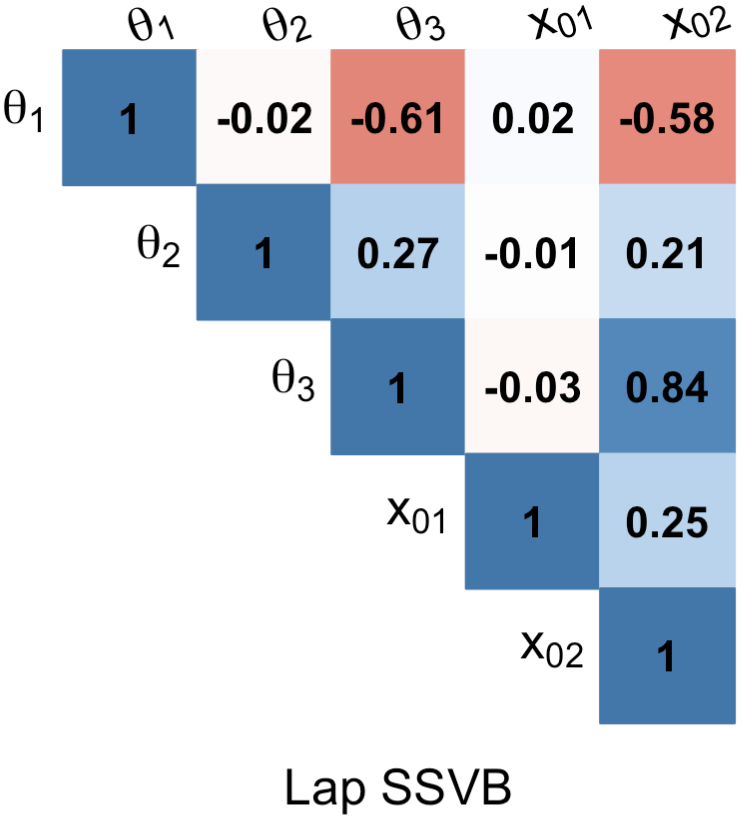}
		\subcaption*{}
	\end{subfigure}
	\begin{subfigure}{0.24\linewidth}
		\includegraphics[width=\linewidth]{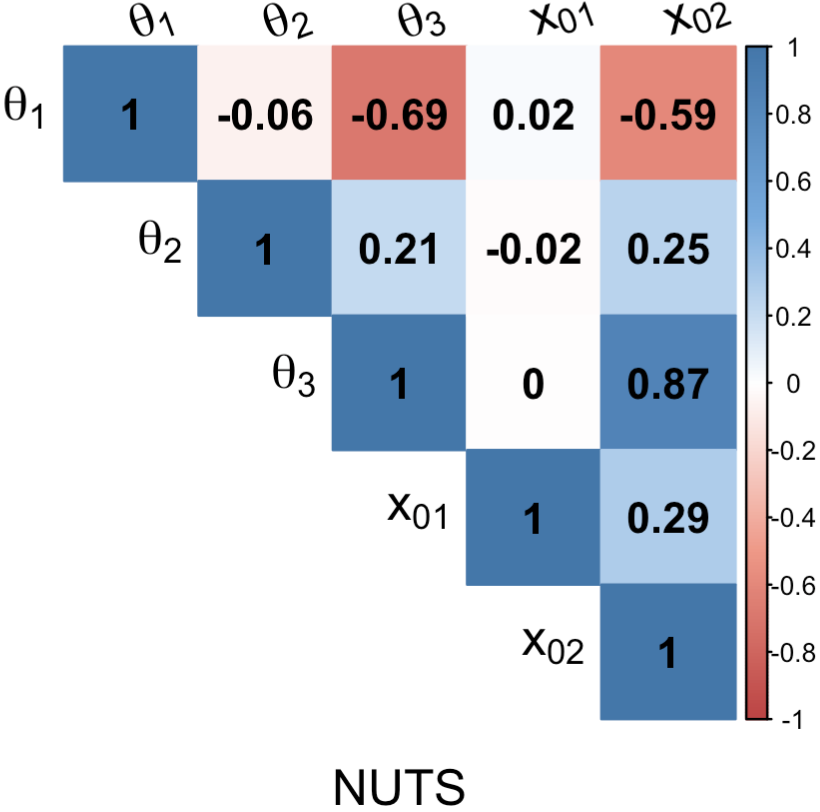}
		\subcaption*{}
	\end{subfigure}
	\ 
	\begin{subfigure}{0.24\linewidth}
		\includegraphics[width=\linewidth]{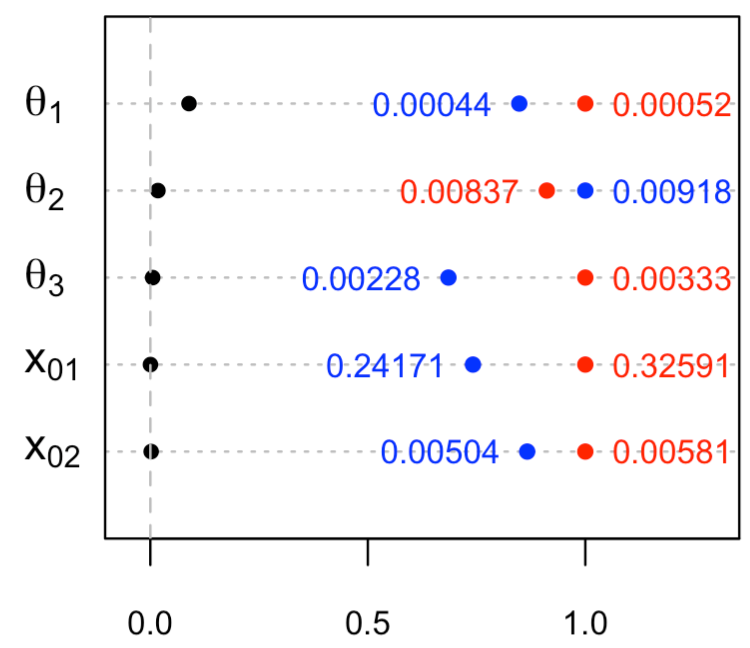}
		\subcaption*{}
	\end{subfigure} \\
	\hrulefill
	\newline
	\begin{subfigure}{0.24\linewidth}
		\includegraphics[width=\linewidth]{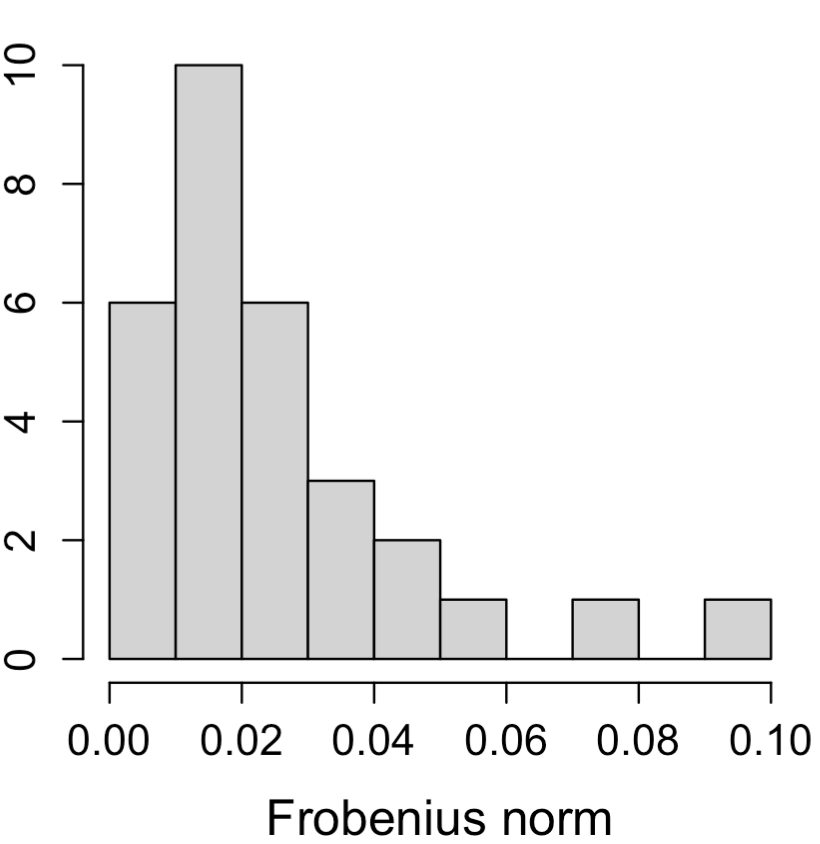}
		\subcaption{Lap Org - DRAM}
	\end{subfigure}
	\begin{subfigure}{0.215\linewidth}
		\includegraphics[width=\linewidth]{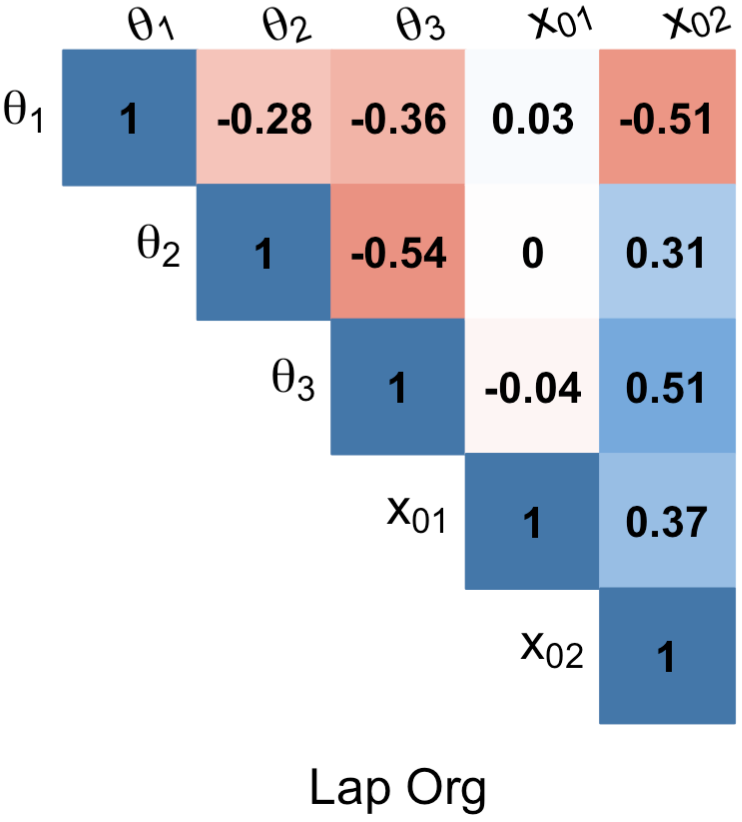}
		\subcaption*{}
	\end{subfigure}
	\begin{subfigure}{0.24\linewidth}
		\includegraphics[width=\linewidth]{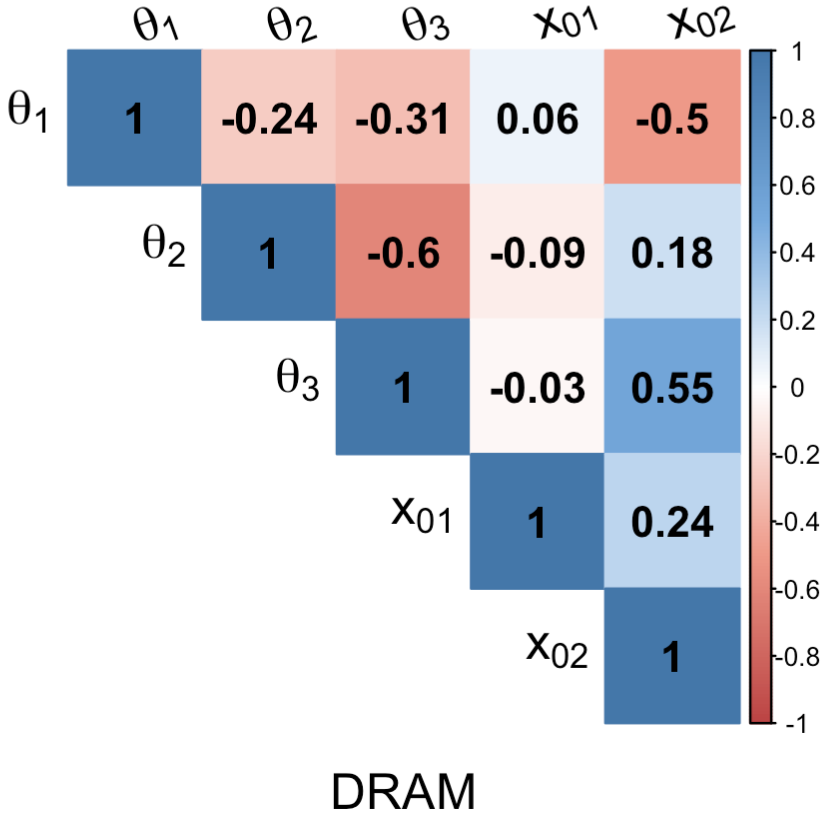}
		\subcaption*{}
	\end{subfigure}
	\ 
	\begin{subfigure}{0.24\linewidth}
		\includegraphics[width=\linewidth]{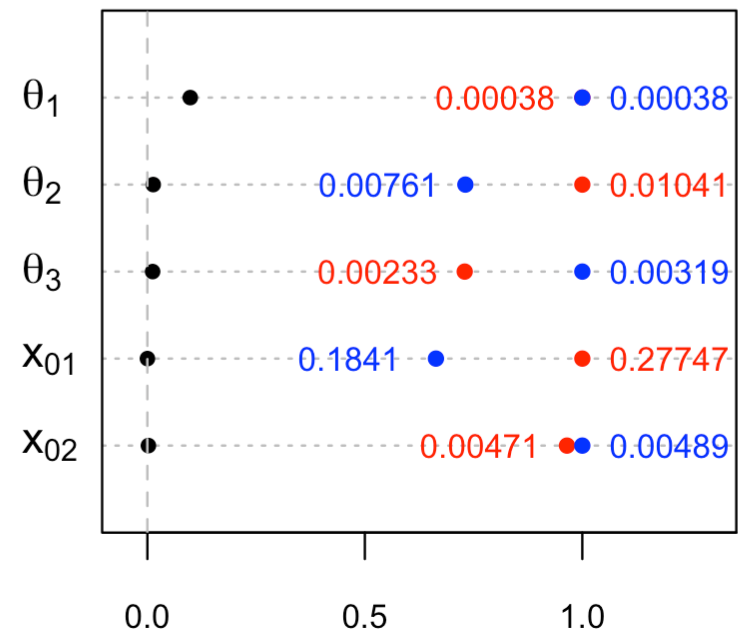}
		\subcaption*{}
	\end{subfigure}\\
	\hrulefill
	\newline
	\begin{subfigure}{0.24\linewidth}
		\includegraphics[width=\linewidth]{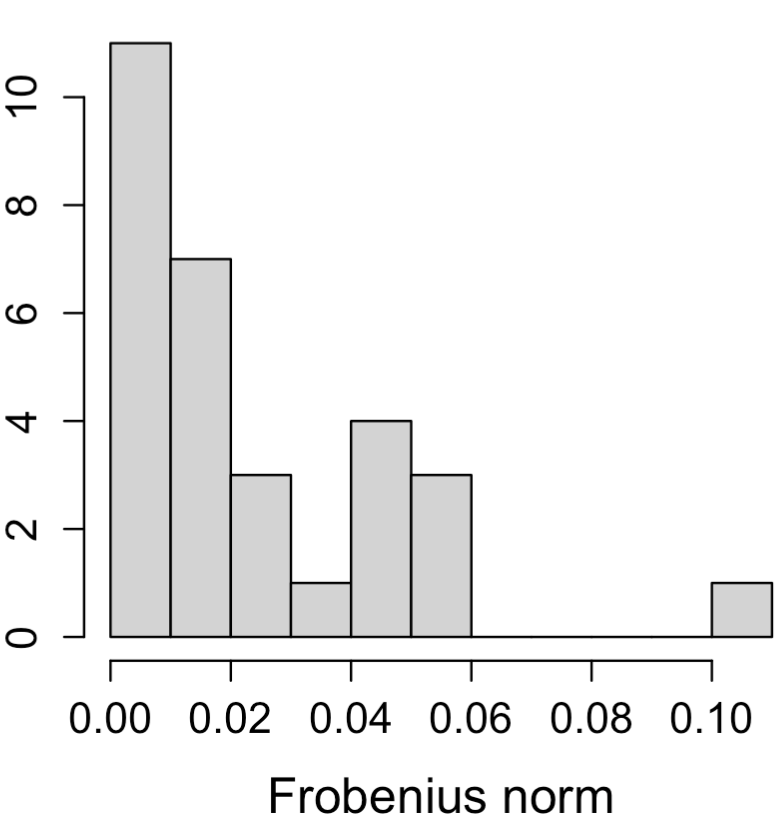}
		\subcaption{Lap Org - NUTS}
	\end{subfigure}
	\begin{subfigure}{0.215\linewidth}
		\includegraphics[width=\linewidth]{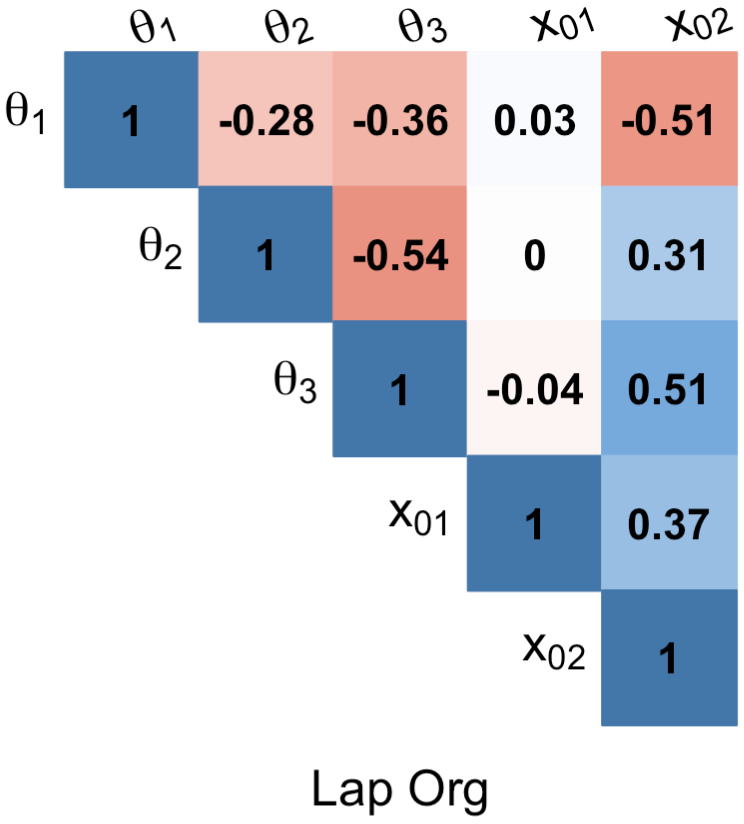}
		\subcaption*{}
	\end{subfigure}
	\begin{subfigure}{0.24\linewidth}
		\includegraphics[width=\linewidth]{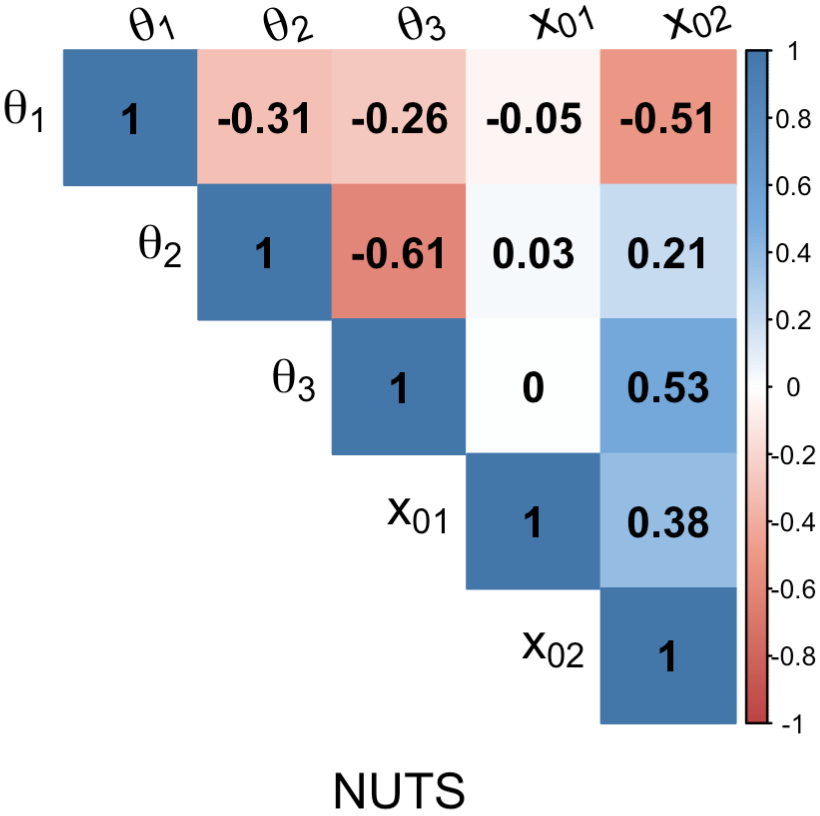}
		\subcaption*{}
	\end{subfigure}
	\ 
	\begin{subfigure}{0.24\linewidth}
		\includegraphics[width=\linewidth]{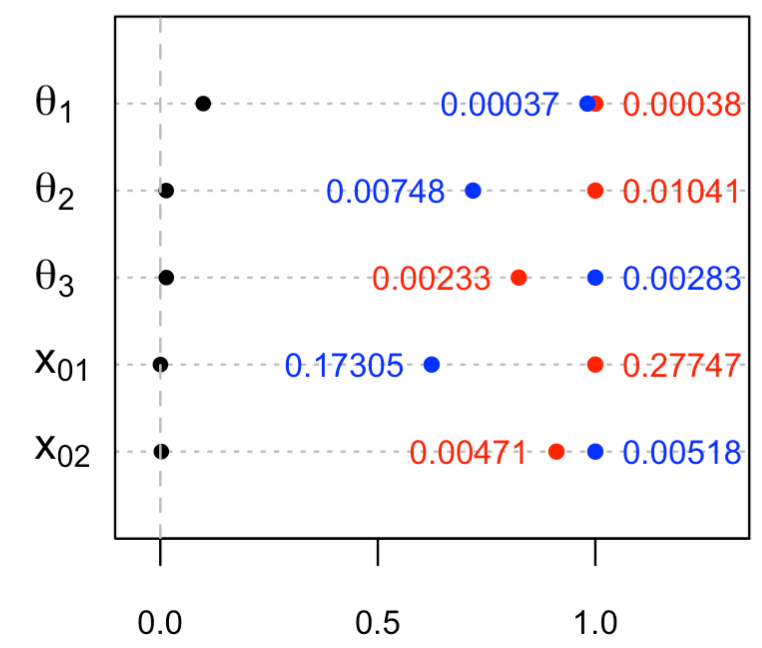}
		\subcaption*{}
	\end{subfigure}
	\caption{Repeat experiment results on the FitzHugh-Nagumo model.  For each pair, the first figure shows the histogram of the Frobenius norm between the two covariance matrices. The results of the case with the largest difference (norm) are presented together. In the variance comparison on the far right, black indicates the mean-field assumption, red indicates the Laplace methods, and blue indicates the MCMC methods. }
	\label{fig:FNrepeated}
\end{figure}

A repeat experiment was conducted to confirm stability. We obtained the covariance matrices in the above four ways for 30 different data sets. These data sets also correspond to indices 1 $\sim$ 30 among the data sets used for the simulation study in \citeasnoun{yang2021variational}. The Frobenius norm 
$$ ||A-B||_F = \sqrt{\sum_{i=1}^n\sum_{j=1}^n|a_{ij}-b_{ij}|^2} $$
was used to measure the similarity between matrices. Figure~\ref{fig:FNrepeated} shows the histogram of the Frobenius norm between the two covariances obtained by the Laplace method and the MCMC method. Among the 30, the results of the case with the largest difference are also shown. The correlation structures were quite similar, even in the most different cases. The magnitude of variance was also much closer to the MCMC methods than under the mean-field assumption. The Laplace approximation method seems stable in improving the posterior covariance.

\newpage
\subsection{Lorenz-96 model}
The Lorenz-96 model, also chosen in \citeasnoun{yang2021variational},
$$  \frac{dX_j}{dt} = \theta_{1j}(X_{j+1}-X_{j-2})X_{j-1}-\theta_{2j}X_j+\theta_{3j},\qquad\text{for }\ j=1,\  \dots,\ p,$$
is a toy model whose size can be adjusted by selecting the number of variables $p$ ($ \geq3 $). In this paper, the experiment was conducted with the case of 4 variables and 12 parameters, considering that the DRAM and NUTS are much slower to run in the larger models. From the true parameter $(\btheta_{1j},\btheta_{2j},\btheta_{3j})=(1,1,8)$ for all $j=1,2,3,4$ and $\bfx_0=(1,8,4,3)^T$, a data set was generated along the 51 equidistant time points $t_0=0,\ t_1=0.1,\ \dots,\ t_{50}=5 $ with the error variance $1/\lambda=1$.

For the DRAM algorithm, we got a chain of 1,000 by thinning at every 200th iteration from 200,000 iterations after the burn-in of 100,000. The NUTS algorithm was implemented using the multi-core option on the \textbf{rstan} package for high speed. A total of 4 chains were obtained, each of 250 iterations by thinning at every 10th from 2,500 after 500 burn-in, providing the resulting samples of 1,000 iterations. The SSVB was run with the tuning parameter
$\tau=0.1^4$ and the step size $m =2 $. The step size of 2 means that each observation time interval in (\ref{SSM}) is divided into 2 subintervals and the approximating function $\bfg(\cdot)$ is applied over the 2 subintervals repeatedly. As $m\ra\infty$ and $\tau\ra0$, the state-space model approaches to the original ODE model. See \citeasnoun{yang2021variational}.

\begin{figure}[b!]
	\centering
	\includegraphics[width=0.9\linewidth]{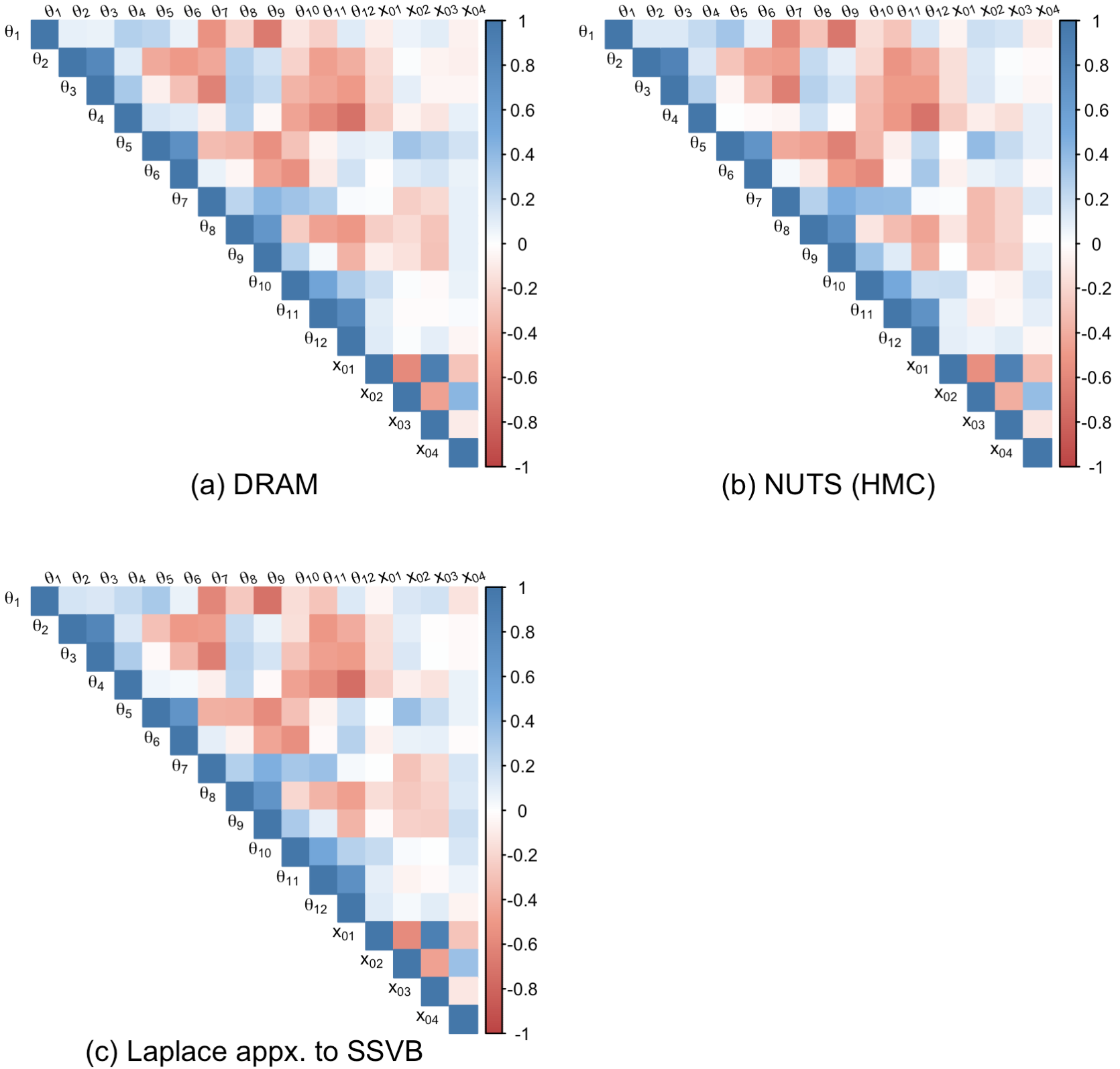}
	\caption{Posterior correlation structures of the Lorenz-96 model from (a) DRAM, (b) NUTS (HMC), (c) SSVB with Laplace approximation.}
	\label{fig:LorenzCorr}
\end{figure}

Unlike the FitzHugh-Nagumo model above, the Laplace approximation to the original ODE model was excluded in this experiment. In the computation, some correlation coefficients were outside of the range $ [-1 ,1] $. We believe that in the calculation of the first and second derivatives of an ODE solution $\bfx(t;\btheta, \bfx_0)$ with respect to $(\btheta^T,\bfx_0^T)^T$, numeric errors seem to be accumulated. These values are obtained as a solution of the extended ODE system with more than $ \half\ p(p+q)(p+q+1) = 544 $ variables (see Appendix \ref{sensitivity}). In addition to the high sensitivity of the ODE system itself, such a large scale of the extended ODE system can be expected to cause huge accumulated errors in numerical solutions. The magnitude of the variance estimates, which is irregularly far from the results of the other algorithms, also supported the guess.

Figure~\ref{fig:LorenzCorr} shows the correlation structure estimates for the Lorenz-96 model. As in the case of the FitzHugh-Nagumo model, the Laplace approximation applied to the SSVB provided a structure quite similar to the MCMC methods. Certainly, the modified one is a better estimate than just an identity matrix.

\begin{figure}[b!]
	\centering
	\includegraphics[width=0.7\linewidth]{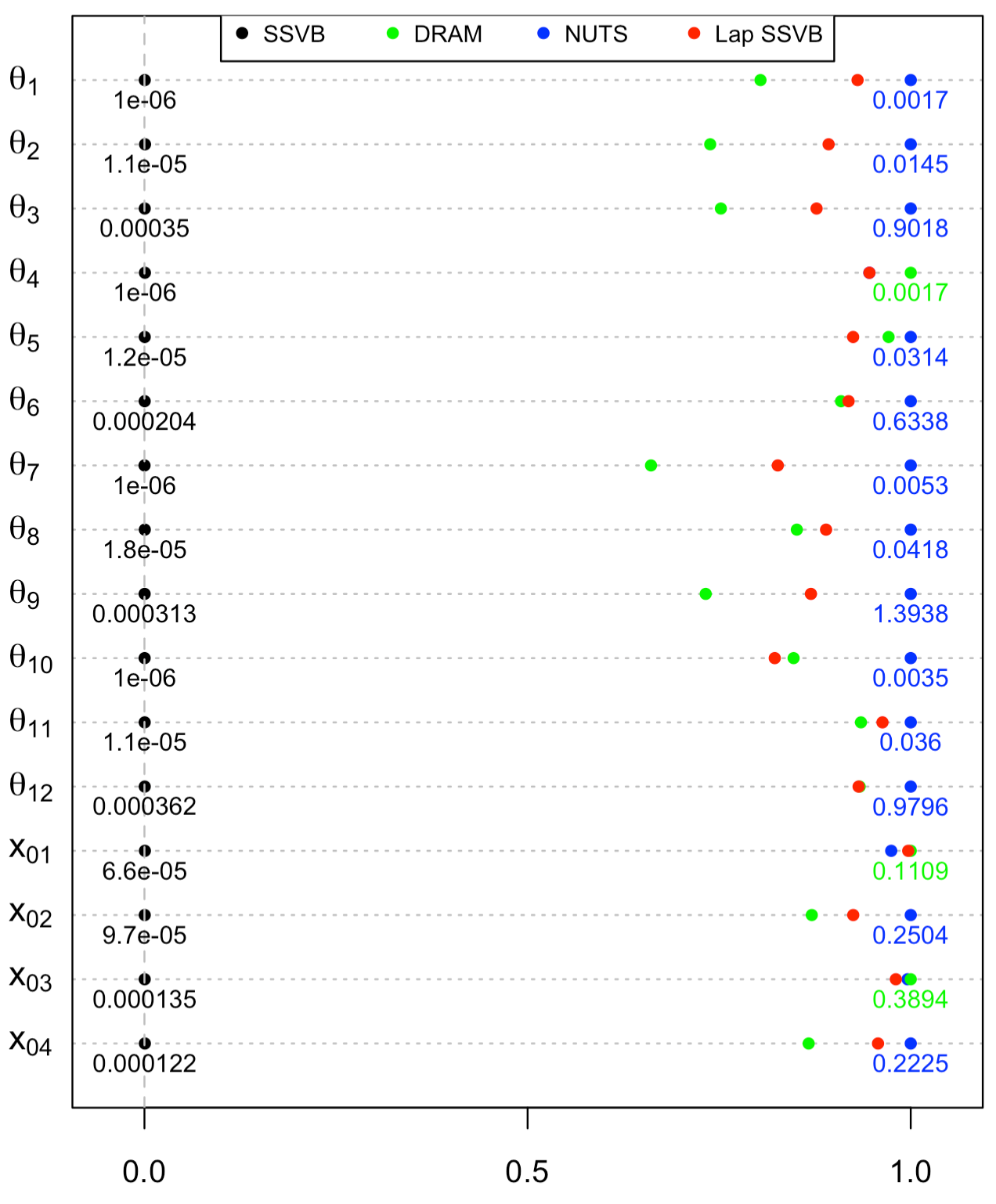}
	\caption{Posterior variance estimates of the Lorenz-96 model. For intuitive visualization of the scale, the estimates relative to the largest one are plotted on a horizontal line for each parameter.}
	\label{fig:LorenzVar}
\end{figure}

The variance estimates for the Lorenz-96 model are plotted in Figure~\ref{fig:LorenzVar}. In this case, as well, the Laplace approximation made the underestimated variance of the SSVB dramatically jump, placing it near the estimates of the DRAM and NUTS. Unlike the FitzHugh-Nagumo model, the tendency to estimate the modified variance larger than the MCMC ones was not evident in this case. This seems to be the effect of the step size of 2, making the relaxed model closer to the original ODE model. Obviously, the Laplace approximation looks very meaningful in terms of the correction for the rough scale.

\begin{figure}[t!]
	\centering
	\begin{subfigure}{0.24\linewidth}
		\includegraphics[width=\linewidth]{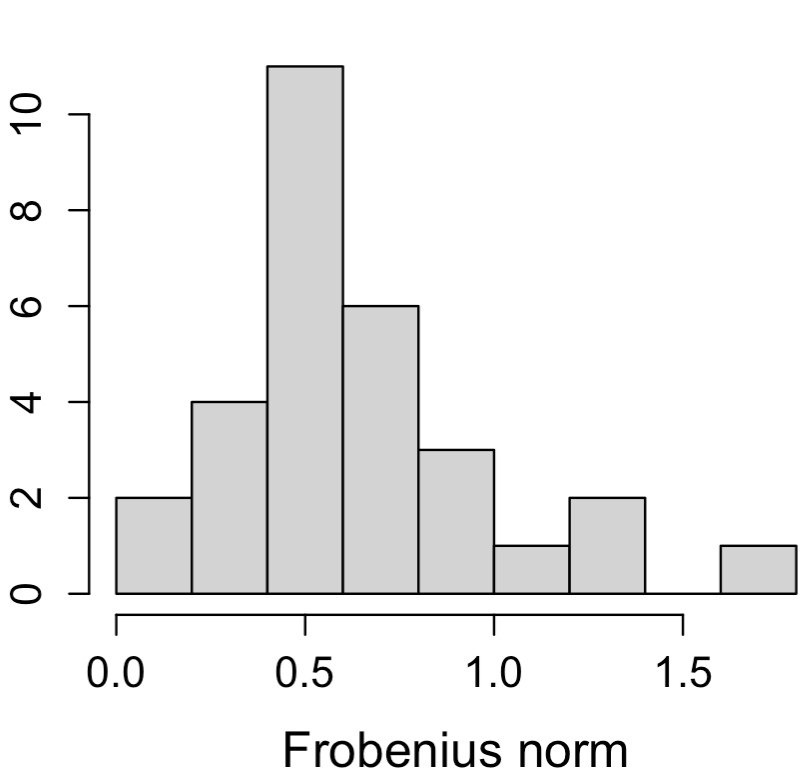} 
		\subcaption{Lap SSVB - DRAM}
	\end{subfigure}
	\begin{subfigure}{0.22\linewidth}
		\includegraphics[width=\linewidth]{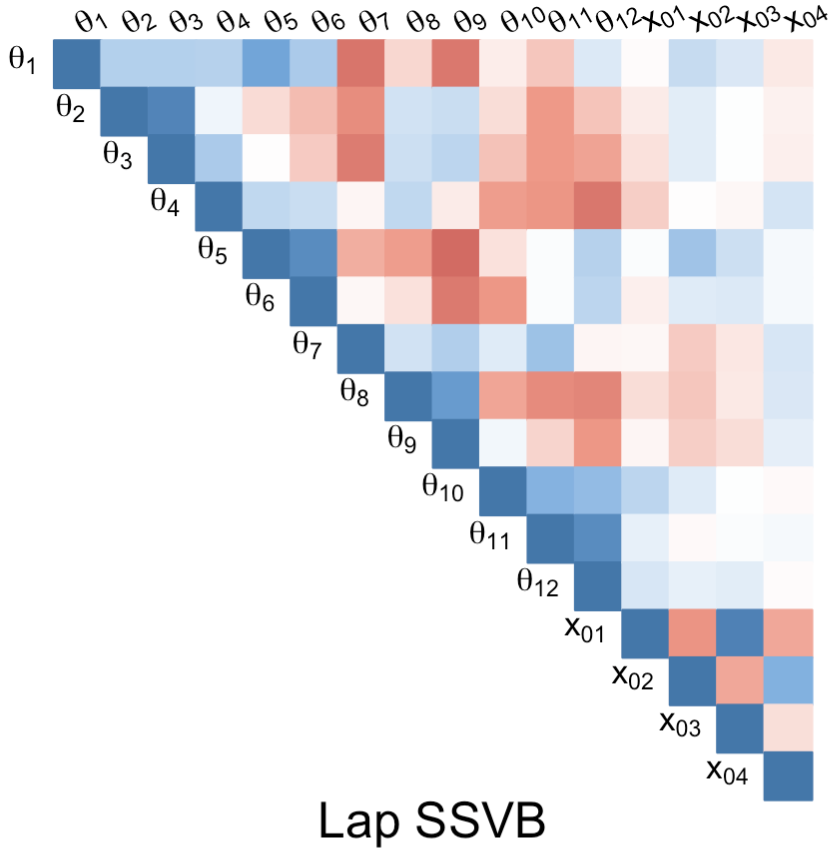}
		\subcaption*{}
	\end{subfigure}
	\begin{subfigure}{0.24\linewidth}
		\includegraphics[width=\linewidth]{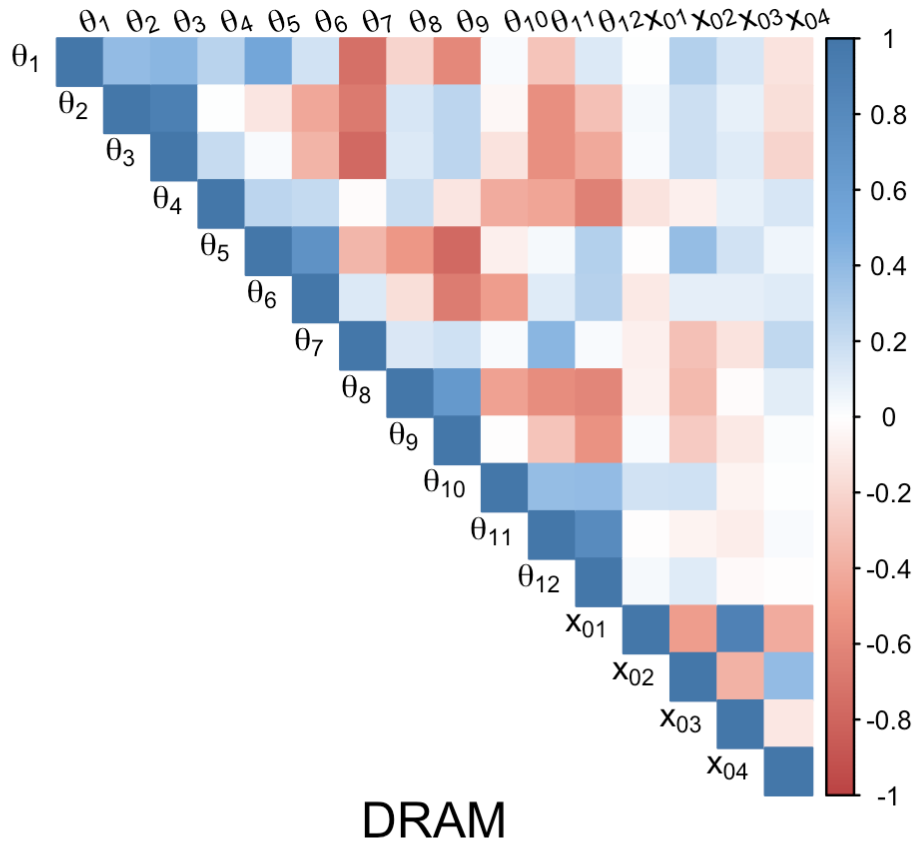}
		\subcaption*{}
	\end{subfigure}
	\ 
	\begin{subfigure}{0.24\linewidth}
		\includegraphics[width=\linewidth]{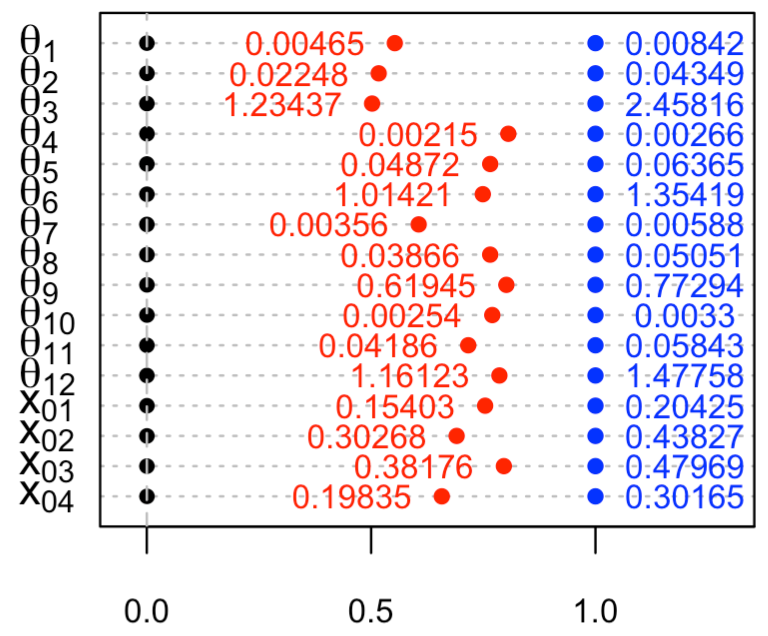}
		\subcaption*{}
	\end{subfigure}\\
	\hrulefill 
	\newline
	\begin{subfigure}{0.24\linewidth}
		\includegraphics[width=\linewidth]{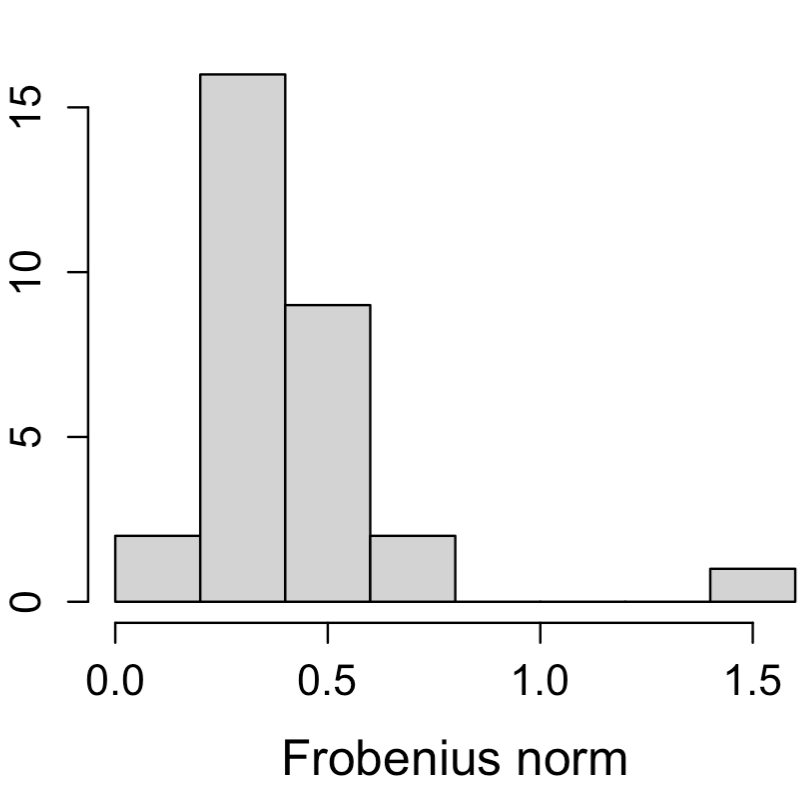}
		\subcaption{Lap SSVB - NUTS}
	\end{subfigure}
	\begin{subfigure}{0.22\linewidth}
		\includegraphics[width=\linewidth]{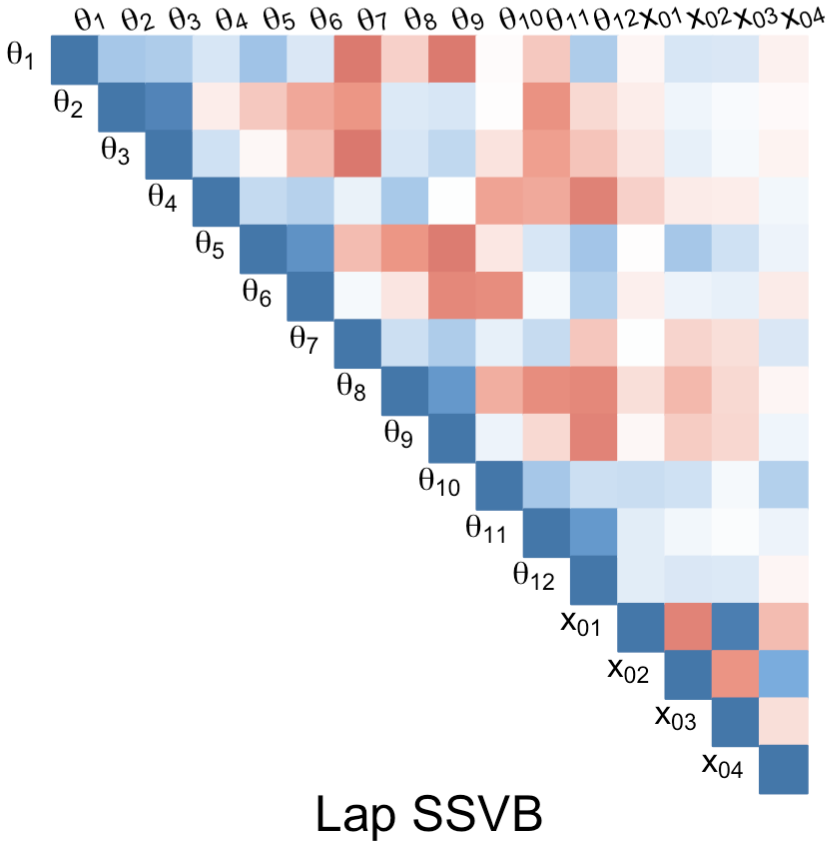}
		\subcaption*{}
	\end{subfigure}
	\begin{subfigure}{0.24\linewidth}
		\includegraphics[width=\linewidth]{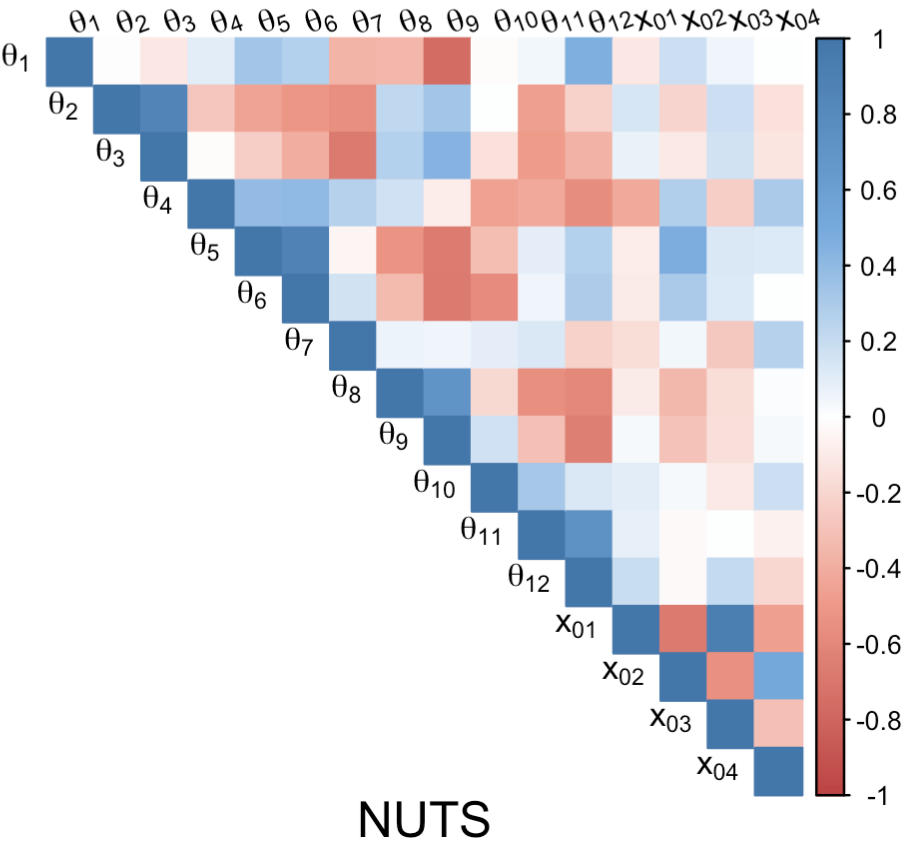}
		\subcaption*{}
	\end{subfigure}
	\ 
	\begin{subfigure}{0.24\linewidth}
		\includegraphics[width=\linewidth]{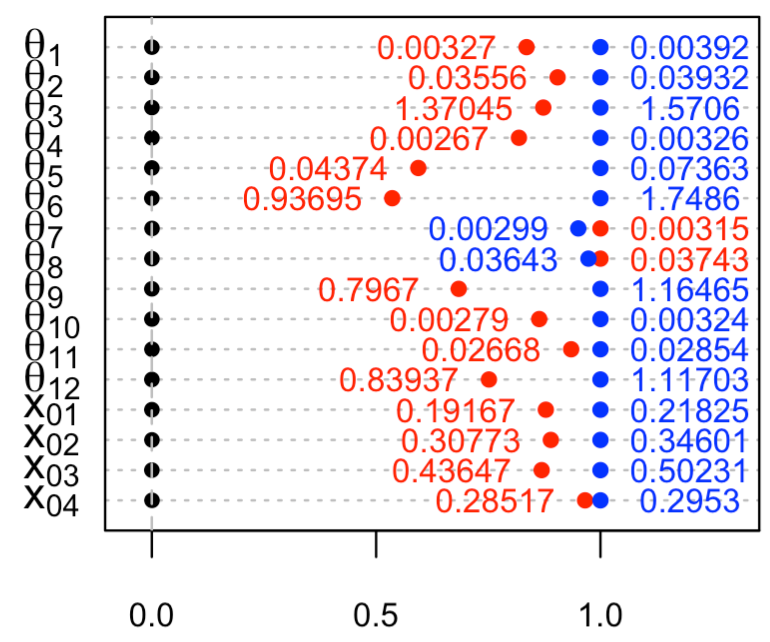}
		\subcaption*{}
	\end{subfigure} 
	\caption{Repeat experiment results on the Lorenz-96 model.  For each pair, the first figure shows the histogram of the Frobenius norm between the two covariance matrices. The results of the case with the largest difference (norm) are presented together. In the variance comparison on the far right, black indicates the mean-field assumption, red indicates the Laplace method, and blue indicates the MCMC methods. }
	\label{fig:L96repeated}
\end{figure}

A repeat experiment was also performed on the Lorenz-96 model.  Similar to the FitzHugh-Nagumo model, Figure~\ref{fig:L96repeated} shows the results for the data sets corresponding to indices 1 to 30 used for the simulation study in \citeasnoun{yang2021variational}. Even in the case with the largest difference, the correlation structures were very similar except for few small correlations. The improvement of the variance magnitude was also evident for that case. The considerable stability was confirmed once again.

\section{Application to real data: COVID-19}
\label{s:Lap_appl}
In \citeasnoun{yang2021variational}, a SIR model with great flexibility,
\begin{align*}
	\frac{dI(t)}{dt}&=\ \frac{\beta(t) I(t)(N-I(t)-R(t))}{N}-\gamma(t) I(t), \\ 
	\frac{dR(t)}{dt}&=\ \gamma(t) I(t), \\
	\text{where }\ \beta(t) &:= \exp\Big\{\sum_i c_{\beta,i}B_{\beta,i}(t)\Big\}, \\
	\gamma(t) &:= \exp\Big\{\sum_i c_{\gamma,i} B_{\gamma,i}(t)\Big\},
\end{align*}
was devised for COVID-19 data fitting. The number of infectious people $I(t)$ and the number of removed people $R(t)$ are modeled with time-varying $\beta(t)$ and $\gamma(t)$ using the cubic B-spline basis functions. With 30 basis functions each, a total of 60 basis coefficients, the ODE parameters, were properly estimated by the SSVB algorithm for the South Korea data. For details, see \citeasnoun{yang2021variational}.

When applying the Laplace approximation to the time-varying SIR model, however,  significantly more parameters than the previous experiments cause numerical problems related to the inverse matrix computation. The straightforward computation following Section~\ref{subsec:lap_to_relaxed} results in the precision matrix $\bfH$ being not positive definite, which also yields the non-positive definite covariance matrix $\bfH^{-1}$. Actually, the resulting covariance matrix included some negative variances. Two strategies were used to correct this.

\begin{figure}[b!]
	\centering
	\begin{subfigure}{0.24\linewidth}
		\includegraphics[width=\linewidth]{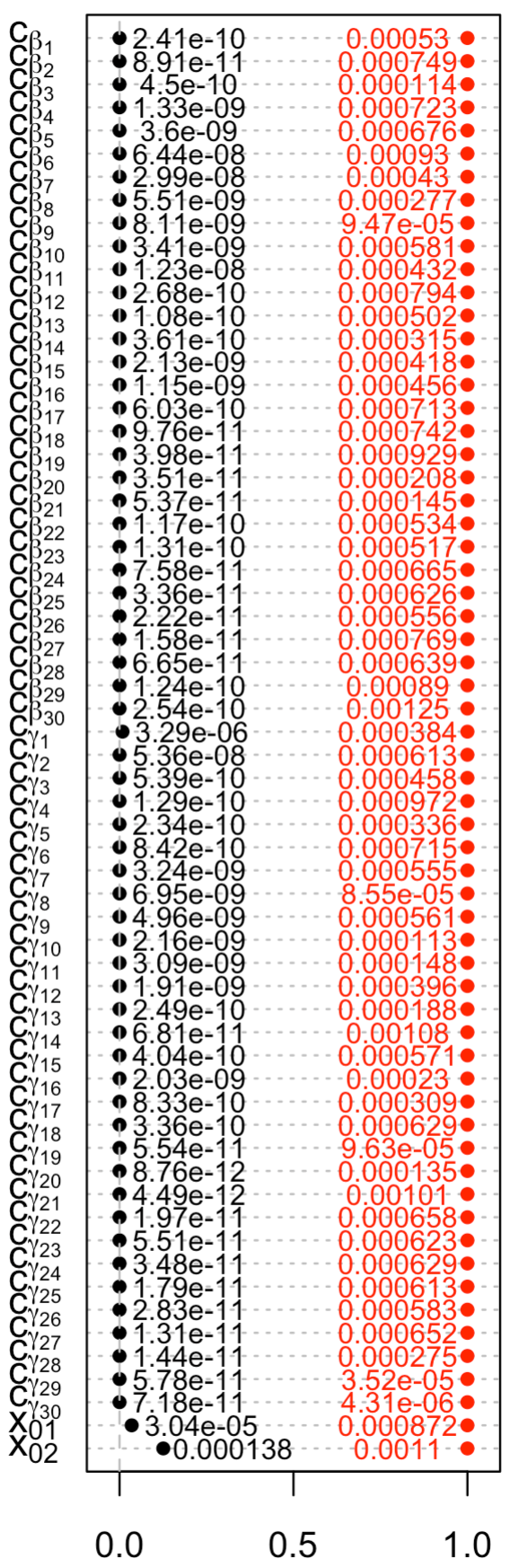}
	\end{subfigure}
	\begin{subfigure}{0.73\linewidth}
		\includegraphics[width=\linewidth]{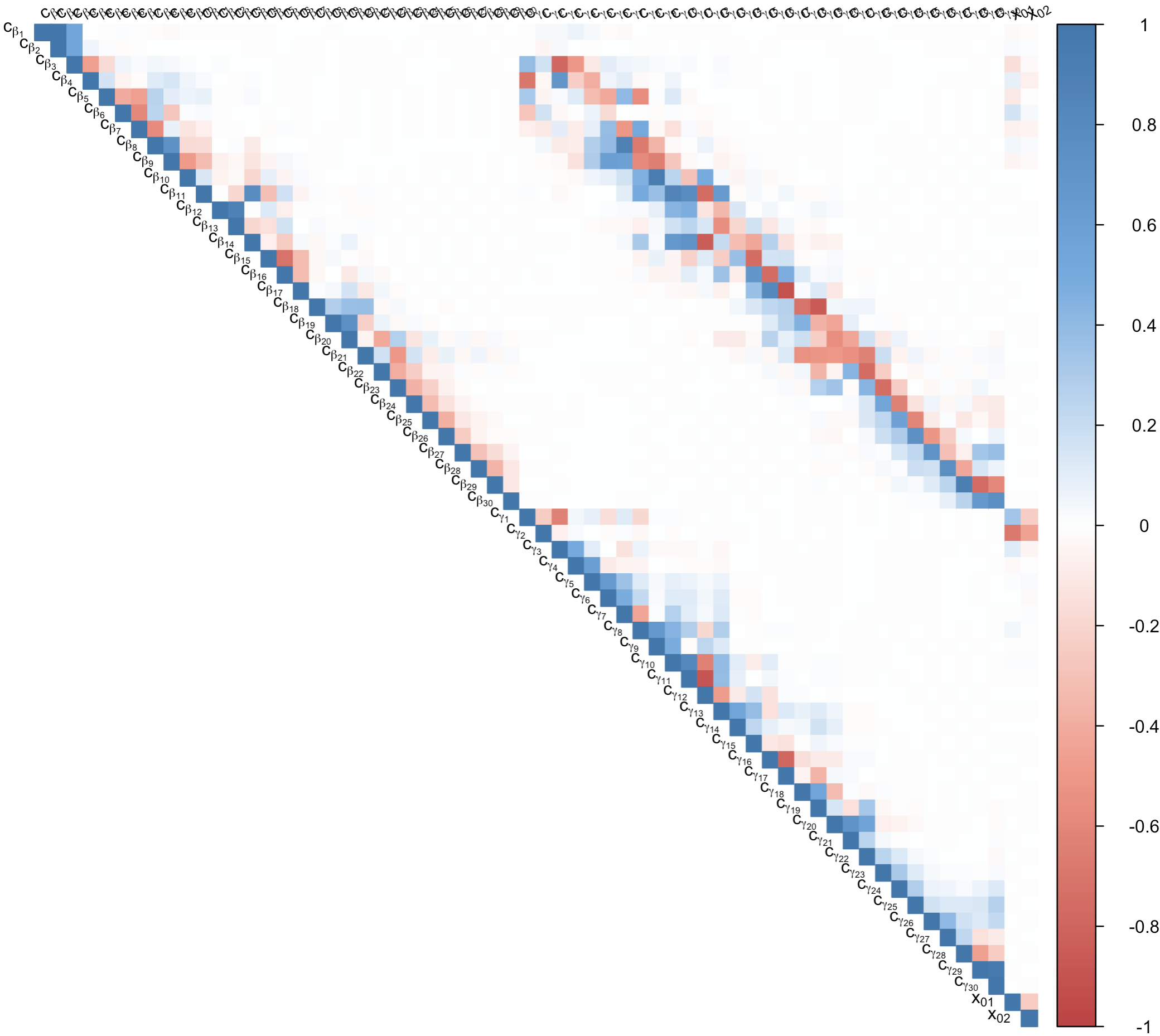}
	\end{subfigure}
	\caption{\textit{Left}: Posterior variance estimates of the time-varying SIR model. Black ones for the SSVB, and red ones for the SSVB with Laplace approximation. For intuitive visualization of the scale, the relative positions are plotted on a horizontal line for each parameter. \textit{Right}: Posterior correlation structure of the time-varying SIR model from the SSVB with Laplace approximation.}
	\label{fig:SIR_corr_var}
\end{figure} 

The first is the inverse of a partitioned matrix. A four partitioned matrix can be inverted by
$$ \begin{bmatrix} 
\bfA &\bfB \\ 
\bfC & \bfD \\ 
\end{bmatrix} ^{-1} = 
\begin{bmatrix} 
(\bfA-\bfB\bfD^{-1}\bfC)^{-1}  & -(\bfA-\bfB\bfD^{-1}\bfC)^{-1} \bfB\bfD^{-1} \\ 
-\bfD^{-1}\bfC(\bfA-\bfB\bfD^{-1}\bfC)^{-1}  & \bfD^{-1}+\bfD^{-1}\bfC(\bfA-\bfB\bfD^{-1}\bfC)^{-1} \bfB\bfD^{-1} \\ 
\end{bmatrix},
$$
when $\bfD$ is nonsingular \cite{ouellette1981schur}. When the precision matrix $\bfH$ is partitioned into the above four blocks with square block $\bfA$ corresponding to the position for $(\btheta,\bfx_0)$, excluding the other state variables $(\bfx_1,\dots,\bfx_n,\lambda)$, the covariance matrix for $(\btheta,\bfx_0)$ can be obtained by inverting $\bfA-\bfB\bfD^{-1}\bfC$. Compared to inverting the whole $\bfH$, the numerical error can be reduced as the dimension of the matrix to be inverted becomes smaller. Indeed, the recovery of the underestimated variances differed greatly depending on whether this strategy was used or not.

The second is to enforce positive definiteness. The resulting $\bfA-\bfB\bfD^{-1}\bfC$ was not positive definite, and neither was its inverse, the covariance matrix. For this case, \textbf{R} library \texttt{Matrix} provides  \texttt{nearPD} function that computes the nearest positive definite matrix to a given matrix, based on \citeasnoun{higham2002computing}. By using the nearest positive definite matrix of $\bfA-\bfB\bfD^{-1}\bfC$, we can obtain a qualified covariance matrix. 

\begin{figure}[b!]
	\centering
	\includegraphics[width=0.6\linewidth]{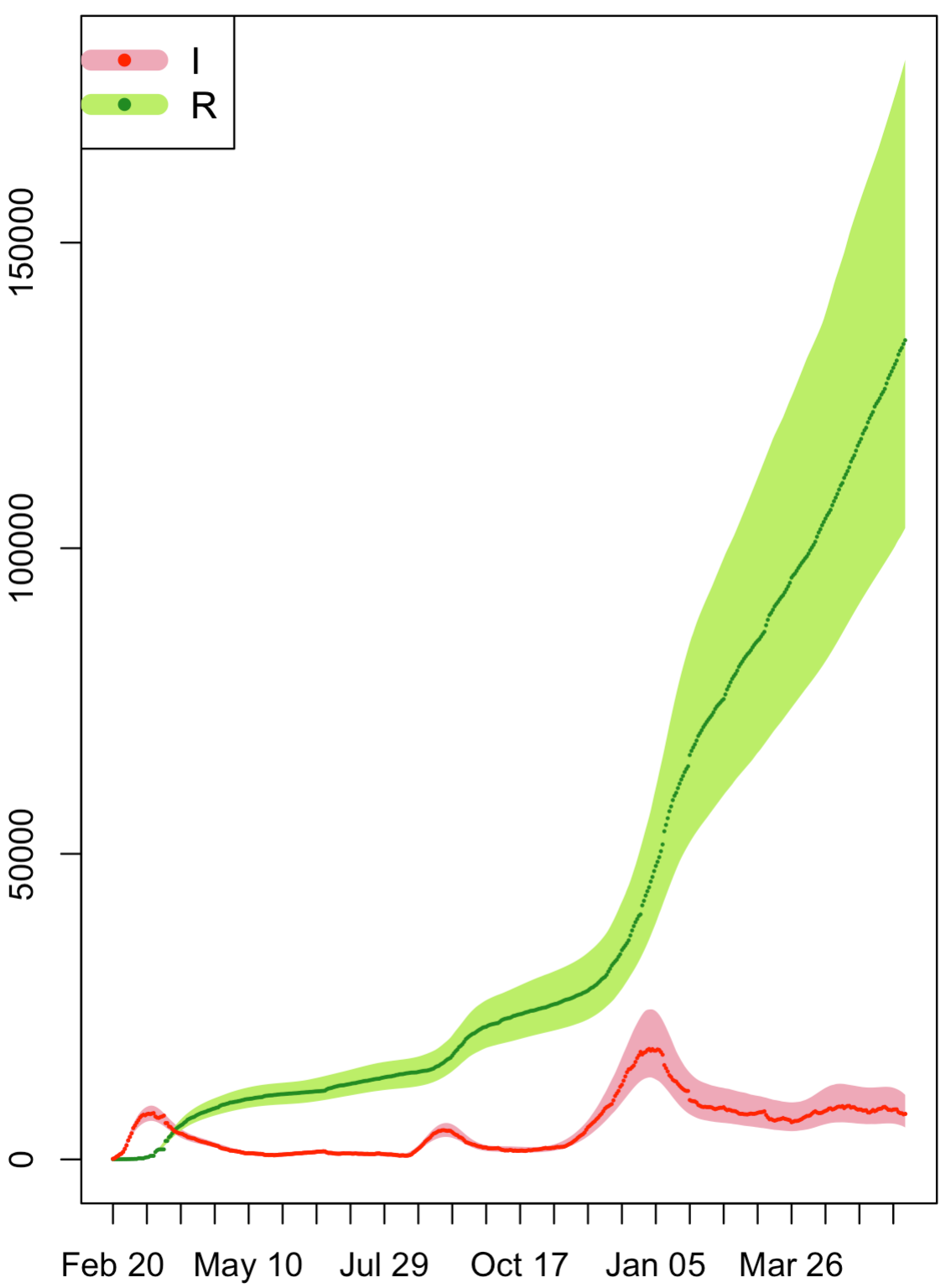}
	\caption{COVID-19 data of Korea with the 95\% posterior credible intervals for the ODE solution curves. The points in red \& green represent the data, and the pink \& light green areas represent the intervals.}
	\label{fig:SIR_sol}
\end{figure}

Figure~\ref{fig:SIR_corr_var} shows the result of applying the Laplace approximation to the time-varying SIR model with the COVID-19 data of Korea. In the left figure, the underestimated variances of the SSVB have jumped significantly by the Laplace approximation. The right figure represents the correlation structure modified by the Laplace approximation. According to the characteristics of B-spline basis functions whose values are non-zero in only specific time intervals, the adjacent basis functions' coefficients showed the more correlated results. Similarly, the coefficients for $\beta(t)$ and $\gamma(t)$ with the same basis function were more correlated with each other.

With a posterior sample $(\btheta^*,\bfx_0^*)$ from the resulting covariance matrix, a posterior sample of the ODE solution curve, $\bfx^*(t;\btheta^*,\bfx_0^*)$, can be obtained. Figure~\ref{fig:SIR_sol} represents the 95\% posterior credible intervals for the ODE solution curves with the data of Korea. A total of 1,000 sample curves were drawn, and the intervals based on the (25th, 975th)-largest values at the observation times are plotted. Due to the sensitivity of the ODE solution, the credible intervals not being wide at the beginning generally widen over time. With the plausible covariance, it has become possible to represent the uncertainty of the regression curve graphically.

\section{Discussion}
\label{s:Lap_disc}
In this paper, we propose to improve the posterior covariance estimation, which is a weakness of the SSVB method, using the Laplace approximation. To confirm how valid these modifications are, the MCMC-based methods were chosen as a reasonable standard on behalf of the true posterior covariance virtually impossible to know. Although MCMC methods do not perform well in ODE parameter estimation itself due to the limitations of the sampling method, they can be manipulated to obtain an ideal chain by selecting the chain starting point close to the true parameters.

The effect of the Laplace approximation seems quite valid. In the experiments, the correlation structure of the SSVB, which is just an identity matrix due to the mean-field assumption, has become very similar to that of the MCMC methods. Furthermore, the underestimated variance, the uncertainty of the inference, also showed a jump in a scale comparable to that of the MCMC methods. Though we cannot say this is the perfect answer, it seems clear that it is a far better estimate than the simple SSVB's one under the mean-field assumption.

For larger ODE models, such as the Lorenz-96 model with 10 variables, the proposed Laplace method can be applied but the results are not included in this paper. It is because the execution time of the MCMC methods, which was set as a standard for determining the improvement, is too long. From the results on the smaller models compared in Section \ref{s:Lap_simul}, it is but inferred that the Laplace approximation could also provide a better covariance estimate in larger ODE models.

As an example, the proposed method was also applied to the time-varying SIR model with 60 ODE parameters for the COVID-19 data of Korea. With some strategies to overcome the numerical problems, we were able to obtain better covariance and use them to represent the posterior credible intervals.

The Laplace approximation is quick to execute, as long as the second derivatives and the mode are given. The success of the Laplace approximation is fundamentally based on excellent parameter estimation of the SSVB used as a mode. By combining the fast and stable parameter estimating method with a good enough density approximation, we can obtain a fairly useful ODE estimation algorithm.

\newpage
\appendix
\section{Appendix : The full second derivatives for the Laplace approximation of the relaxed model }\label{2ndFull}
\begin{align*}
&\frac{\partial^2}{\partial\lambda^2} = \frac{1}{\lambda^2}\left( \frac{p(n+1)}{2}+A_0-1 \right), \\
&\frac{\partial^2}{\partial\btheta\partial\lambda} = \bzero, \\
&\frac{\partial^2}{\partial\bfx_i\partial\lambda} = \bfx_i-\bfy_i \qquad \text{for } i = 0,\dots,n, \\
&\frac{\partial^2}{\partial\btheta\partial\btheta^T} = -\frac{1}{\tau} \sum_{i=1}^{n} \left[\sum_{j=1}^p (x_{ij}-g_j(\bfx_{i-1},\btheta))\cdot \frac{\partial^2 g_j(\bfx_{i-1},\btheta)}{\partial\btheta\partial\btheta^T} - \bfJ_{\bfg\text{ wrt }\btheta}(\bfx_{i-1},\btheta)^T\bfJ_{\bfg\text{ wrt }\btheta}(\bfx_{i-1},\btheta) \right], \\
&\frac{\partial^2}{\partial\bfx_0\partial\btheta^T} = -\frac{1}{\tau} \left[ \sum_{j=1}^p (x_{1j}-g_j(\bfx_{0},\btheta))\cdot \frac{\partial^2 g_j(\bfx_{0},\btheta)}{\partial\bfx\partial\btheta^T} - \bfJ_{\bfg\text{ wrt }\bfx}(\bfx_{0},\btheta)^T\bfJ_{\bfg\text{ wrt }\btheta}(\bfx_{0},\btheta) \right], \\
&\text{ for }i=1,\dots,n-1,  \\
&\frac{\partial^2}{\partial\bfx_i\partial\btheta^T} = -\frac{1}{\tau} \left[ \sum_{j=1}^p (x_{i+1,j}-g_j(\bfx_{i},\btheta))\cdot \frac{\partial^2 g_j(\bfx_{i},\btheta)}{\partial\bfx\partial\btheta^T} - \bfJ_{\bfg\text{ wrt }\bfx}(\bfx_{i},\btheta)^T\bfJ_{\bfg\text{ wrt }\btheta}(\bfx_{i},\btheta) + \bfJ_{\bfg\text{ wrt }\btheta}(\bfx_{i-1},\btheta) \right], \\
&\frac{\partial^2}{\partial\bfx_n\partial\btheta^T} = -\frac{1}{\tau} \bfJ_{\bfg\text{ wrt }\btheta}(\bfx_{n-1},\btheta), \\
&\frac{\partial^2}{\partial\bfx_0\partial\bfx_0^T} = -\frac{1}{\tau} \left[ \sum_{j=1}^p (x_{1j}-g_j(\bfx_{0},\btheta))\cdot \frac{\partial^2 g_j(\bfx_{0},\btheta)}{\partial\bfx\partial\bfx^T} - \bfJ_{\bfg\text{ wrt }\bfx}(\bfx_{0},\btheta)^T\bfJ_{\bfg\text{ wrt }\bfx}(\bfx_{0},\btheta) \right] + \lambda\bfI_p, \\
&\frac{\partial^2}{\partial\bfx_1\partial\bfx_0^T} = -\frac{1}{\tau} \bfJ_{\bfg\text{ wrt }\bfx}(\bfx_0,\btheta), \\
&\text{ for } i=1,\dots,n-1,  \\
&\frac{\partial^2}{\partial\bfx_i\partial\bfx_i^T} = -\frac{1}{\tau} \left[ \sum_{j=1}^p (x_{i+1,j}-g_j(\bfx_{i},\btheta))\cdot \frac{\partial^2 g_j(\bfx_{i},\btheta)}{\partial\bfx\partial\bfx^T} - \bfJ_{\bfg\text{ wrt }\bfx}(\bfx_{i},\btheta)^T\bfJ_{\bfg\text{ wrt }\bfx}(\bfx_{i},\btheta) \right] +\left(\frac{1}{\tau}+\lambda\right) \bfI_p, \\
&\frac{\partial^2}{\partial\bfx_{i+1}\partial\bfx_i^T} = -\frac{1}{\tau} \bfJ_{\bfg\text{ wrt }\bfx}(\bfx_i,\btheta), \\
&\frac{\partial^2}{\partial\bfx_n\partial\bfx_n^T} = \left(\frac{1}{\tau}+\lambda\right) \bfI_p.
\end{align*}

\section{Appendix : Hessian matrix of the 4th Runge-Kutta method}\label{Hessian}

\subsection{Basic chain rule for Hessian}
If $y=f(\bfu)$ and $ \bfu=g(\bfx) $, then the second derivative of $ f\circ g $ is:
$$ \frac{\partial^2 y}{\partial x_i \partial x_j}=\sum_k\left(\frac{\partial y}{\partial u_k} \frac{\partial^2u_k}{\partial x_i \partial x_j} \right)+\sum_{k,\ell}\left(\frac{\partial^2y}{\partial u_k\partial u_\ell}\frac{\partial u_k}{\partial x_i}\frac{\partial u_\ell}{\partial x_j}\right). $$

\subsection{The case of step size $m=1$}
\begin{align*}
\bfg(\bfx,t,\btheta)=&\ \bfx+\frac{1}{6}(K_{1}+2K_{2}+2K_{3}+K_{4}),\\
K_{1}=&\ h\cdot \bff(\bfx,t;\btheta),\\
K_{2}=&\ h\cdot \bff(\bfx+\frac{1}{2}K_{1},t+\frac{1}{2}h;\btheta),\\
K_{3}=&\ h\cdot \bff(\bfx+\frac{1}{2}K_{2},t+\frac{1}{2}h;\btheta),\\
K_{4}=&\ h\cdot \bff(\bfx+K_{3},t+h;\btheta).
\end{align*}
Let $ \bfu = (\bfx^T, \btheta^T)^T $ and the Hessian matrix $ \bfH_{f_j}=\frac{\partial^2 f_j}{\partial\bfu\partial\bfu^T} $ for $j = 1,\dots,p$ are given.
$$ \frac{\partial^2 g_j}{\partial\bfu\partial\bfu^T} = \frac{1}{6}\left(\frac{\partial^2 K_{1j}}{\partial\bfu\partial\bfu^T}+2\frac{\partial^2 K_{2j}}{\partial\bfu\partial\bfu^T}+2\frac{\partial^2 K_{3j}}{\partial\bfu\partial\bfu^T}+
\frac{\partial^2 K_{4j}}{\partial\bfu\partial\bfu^T}\right).  $$

\noindent For $K_1$,
$$ \frac{\partial^2 K_{1j}}{\partial\bfu\partial\bfu^T} = h \cdot \frac{\partial^2 f_j}{\partial\bfu\partial\bfu^T}. $$ 

\begin{align*}
	\text{For } K_2  =\ & h\cdot \bff(\bfv)\text{ when }\bfv = \begin{bmatrix} \bfx+\frac{1}{2}K_{1} \\ \btheta \end{bmatrix} \text{ and } \bfJ_{\bfv\text{ wrt }\bfu} =  \left[
	\begin{array}{c|c}
	\bfI_p+\half \bfJ_{K_1\text{ wrt }\bfx} & \half\bfJ_{K_1\text{ wrt }\btheta} \\
	\hline
	\bfO_{q\times p} & \bfI_q
	\end{array}
	\right], \\
	\frac{\partial^2 K_{2j}}{\partial\bfu\partial\bfu^T} &= h \sum_k \left(\frac{\partial f_j}{\partial v_k}\frac{\partial^2 v_k}{\partial\bfu\partial\bfu^T}\right) + h\cdot \bfJ_{\bfv\text{ wrt }\bfu}^T \left[ \frac{\partial^2 f_j}{\partial\bfv\partial\bfv^T}\right] \bfJ_{\bfv\text{ wrt }\bfu} \\
	&= \half h \sum_{k=1}^p \left\{\bfJ_{\bff\text{ wrt }\bfx}\left(\bfx+\frac{1}{2}K_{1},t+\frac{1}{2}h;\btheta\right)\right\}_{jk} \cdot  \left[\frac{\partial^2 K_{1k}}{\partial\bfu\partial\bfu^T}\right] \\
	&\qquad  + h\cdot \bfJ_{\bfv\text{ wrt }\bfu}^T \ \bfH_{f_j}\left(\bfx+\frac{1}{2}K_{1},t+\frac{1}{2}h;\btheta\right)  \bfJ_{\bfv\text{ wrt }\bfu} .
\end{align*}	

\begin{align*}
\text{For } K_3  =\ & h\cdot \bff(\bfv)\text{ when }\bfv = \begin{bmatrix} \bfx+\frac{1}{2}K_{2} \\ \btheta \end{bmatrix} \text{ and } \bfJ_{\bfv\text{ wrt }\bfu} =  \left[
\begin{array}{c|c}
\bfI_p+\half \bfJ_{K_2\text{ wrt }\bfx} & \half\bfJ_{K_2\text{ wrt }\btheta} \\
\hline
\bfO_{q\times p} & \bfI_q
\end{array}
\right], \\
\frac{\partial^2 K_{3j}}{\partial\bfu\partial\bfu^T} &= h \sum_k \left(\frac{\partial f_j}{\partial v_k}\frac{\partial^2 v_k}{\partial\bfu\partial\bfu^T}\right) + h\cdot \bfJ_{\bfv\text{ wrt }\bfu}^T \left[ \frac{\partial^2 f_j}{\partial\bfv\partial\bfv^T}\right] \bfJ_{\bfv\text{ wrt }\bfu} \\
&= \half h \sum_{k=1}^p \left\{\bfJ_{\bff\text{ wrt }\bfx}\left(\bfx+\frac{1}{2}K_{2},t+\frac{1}{2}h;\btheta\right)\right\}_{jk} \cdot  \left[\frac{\partial^2 K_{2k}}{\partial\bfu\partial\bfu^T}\right] \\
&\qquad  + h\cdot \bfJ_{\bfv\text{ wrt }\bfu}^T \ \bfH_{f_j}\left(\bfx+\frac{1}{2}K_{2},t+\frac{1}{2}h;\btheta\right)  \bfJ_{\bfv\text{ wrt }\bfu} .
\end{align*}	

\begin{align*}
\text{For } K_4  =\ & h\cdot \bff(\bfv)\text{ when }\bfv = \begin{bmatrix} \bfx+K_{3} \\ \btheta \end{bmatrix} \text{ and } \bfJ_{\bfv\text{ wrt }\bfu} =  \left[
\begin{array}{c|c}
\bfI_p+ \bfJ_{K_3\text{ wrt }\bfx} & \bfJ_{K_3\text{ wrt }\btheta} \\
\hline
\bfO_{q\times p} & \bfI_q
\end{array}
\right], \\
\frac{\partial^2 K_{4j}}{\partial\bfu\partial\bfu^T} &= h \sum_k \left(\frac{\partial f_j}{\partial v_k}\frac{\partial^2 v_k}{\partial\bfu\partial\bfu^T}\right) + h\cdot \bfJ_{\bfv\text{ wrt }\bfu}^T \left[ \frac{\partial^2 f_j}{\partial\bfv\partial\bfv^T}\right] \bfJ_{\bfv\text{ wrt }\bfu} \\
&=  h \sum_{k=1}^p \left\{\bfJ_{\bff\text{ wrt }\bfx}\left(\bfx+K_{3},t+h;\btheta\right)\right\}_{jk} \cdot  \left[\frac{\partial^2 K_{3k}}{\partial\bfu\partial\bfu^T}\right] \\
&\qquad  + h\cdot \bfJ_{\bfv\text{ wrt }\bfu}^T \ \bfH_{f_j}\left(\bfx+K_{3},t+h;\btheta\right)  \bfJ_{\bfv\text{ wrt }\bfu} .
\end{align*}

\subsection{The case of step size $m\geq2$}
First, without considering the step size $m$, we can define some iterative functions like:
\begin{align*}
\bfg^{(2)}(\bfx,t,\btheta)=&\ \bfg(\bfg(\bfx,t,\btheta),t+h,\btheta)\\
\bfg^{(3)}(\bfx,t,\btheta)=&\ \bfg(\bfg^{(2)}(\bfx,t,\btheta),t+2h,\btheta)\\ &\vdots \\
\bfg^{(m)}(\bfx,t,\btheta)=&\ \bfg\left(\bfg^{(m-1)}(\bfx,t,\btheta),t+(m-1)h,\btheta\right).	 
\end{align*}

Then, the Jacobian matrices with respect to $\bfx$ of the above functions can be recursively computed from $\bfH_{g_j}(\cdot)$ above as follows: 

\begin{align*}
\text{For } \bfg^{(2)}  =\ & \bfg(\bfv, t+h)\text{ when }\bfv = \begin{bmatrix} \bfg(\bfx,t,\btheta) \\ \btheta \end{bmatrix} \text{ and } \bfJ_{\bfv\text{ wrt }\bfu} =  \left[
\begin{array}{c|c}
 \bfJ_{\bfg\text{ wrt }\bfx} & \bfJ_{\bfg\text{ wrt }\btheta} \\
\hline
\bfO_{q\times p} & \bfI_q
\end{array}
\right], \\
\frac{\partial^2 g_j^{(2)}}{\partial\bfu\partial\bfu^T} &=  \sum_k \left(\frac{\partial g_j}{\partial v_k}\frac{\partial^2 v_k}{\partial\bfu\partial\bfu^T}\right) +  \bfJ_{\bfv\text{ wrt }\bfu}^T \left[ \frac{\partial^2 g_j}{\partial\bfv\partial\bfv^T}\right] \bfJ_{\bfv\text{ wrt }\bfu} \\
&=   \sum_{k=1}^p \left\{\bfJ_{\bfg\text{ wrt }\bfx}\left(\bfg(\bfx,t,\btheta),t+h;\btheta\right)\right\}_{jk} \cdot  \left[\frac{\partial^2 g_k}{\partial\bfu\partial\bfu^T}\right]   +  \bfJ_{\bfv\text{ wrt }\bfu}^T \ \bfH_{g_j}\left(\bfg(\bfx,t,\btheta),t+h;\btheta\right)  \bfJ_{\bfv\text{ wrt }\bfu} .
\end{align*}	

\begin{align*}
\text{For } \bfg^{(3)}  =\ & \bfg(\bfv, t+2h)\text{ when }\bfv = \begin{bmatrix} \bfg^{(2)}(\bfx,t,\btheta) \\ \btheta \end{bmatrix} \text{ and } \bfJ_{\bfv\text{ wrt }\bfu} =  \left[
\begin{array}{c|c}
\bfJ_{\bfg^{(2)}\text{ wrt }\bfx} & \bfJ_{\bfg^{(2)}\text{ wrt }\btheta} \\
\hline
\bfO_{q\times p} & \bfI_q
\end{array}
\right], \\
\frac{\partial^2 g_j^{(3)}}{\partial\bfu\partial\bfu^T} &=  \sum_k \left(\frac{\partial g_j}{\partial v_k}\frac{\partial^2 v_k}{\partial\bfu\partial\bfu^T}\right) +  \bfJ_{\bfv\text{ wrt }\bfu}^T \left[ \frac{\partial^2 g_j}{\partial\bfv\partial\bfv^T}\right] \bfJ_{\bfv\text{ wrt }\bfu} \\
&=   \sum_{k=1}^p \left\{\bfJ_{\bfg\text{ wrt }\bfx}\left(\bfg^{(2)}(\bfx,t,\btheta),t+2h;\btheta\right)\right\}_{jk} \cdot  \left[\frac{\partial^2 g^{(2)}_k}{\partial\bfu\partial\bfu^T}\right] \\
&\qquad  +  \bfJ_{\bfv\text{ wrt }\bfu}^T \ \bfH_{g_j}\left(\bfg^{(2)}(\bfx,t,\btheta),t+2h;\btheta\right)  \bfJ_{\bfv\text{ wrt }\bfu} .
\end{align*}	

$$ \vdots \qquad\qquad\qquad $$ 

\begin{align*}
\text{For } \bfg^{(m)}  =\ & \bfg(\bfv, t+(m-1)h)\text{ when }\bfv = \begin{bmatrix} \bfg^{(m-1)}(\bfx,t,\btheta) \\ \btheta \end{bmatrix} \text{ and } \bfJ_{\bfv\text{ wrt }\bfu} =  \left[
\begin{array}{c|c}
\bfJ_{\bfg^{(m-1)}\text{ wrt }\bfx} & \bfJ_{\bfg^{(m-1)}\text{ wrt }\btheta} \\
\hline
\bfO_{q\times p} & \bfI_q
\end{array}
\right], \\
\frac{\partial^2 g_j^{(m)}}{\partial\bfu\partial\bfu^T} &=  \sum_k \left(\frac{\partial g_j}{\partial v_k}\frac{\partial^2 v_k}{\partial\bfu\partial\bfu^T}\right) +  \bfJ_{\bfv\text{ wrt }\bfu}^T \left[ \frac{\partial^2 g_j}{\partial\bfv\partial\bfv^T}\right] \bfJ_{\bfv\text{ wrt }\bfu} \\
&=   \sum_{k=1}^p \left\{\bfJ_{\bfg\text{ wrt }\bfx}\left(\bfg^{(m-1)}(\bfx,t,\btheta),t+(m-1)h;\btheta\right)\right\}_{jk} \cdot  \left[\frac{\partial^2 g^{(m-1)}_k}{\partial\bfu\partial\bfu^T}\right] \\
&\qquad  +  \bfJ_{\bfv\text{ wrt }\bfu}^T \ \bfH_{g_j}\left(\bfg^{(m-1)}(\bfx,t,\btheta),t+(m-1)h;\btheta\right)  \bfJ_{\bfv\text{ wrt }\bfu} .
\end{align*}	

Now to return to the main, for the given step size $m\geq2$ in our proposed method, we can use  $ \bfH_{g_j^{(m)}}(\bfx,t,\btheta) $ which is computed with $h/m$ instead of $h$ from the above formulas.

\section{Appendix: Sensitivity analysis of ODE systems}
\label{sensitivity}
\subsection{Jacobian of $\bfx(t; \btheta,\bfx_0)$ with respect to $\btheta$ and $\bfx_0$}
When a $p$-variable ODE system $  \dot{\bfx}(t) = \bff(\bfx(t),t,\btheta) $ is given, or  
$$ \dot{x}_j(t) = f_j(\bfx(t),t,\btheta), \quad j=1,\dots,p, $$
consider the partial derivative of the $j$-th solution curve $x_j(t;\btheta,\bfx_0)$ with respect to the $k$-th parameter $\theta_k$,
$$ Z_{kj}(t) \equiv \frac{\partial x_j(t;\btheta,\bfx_0)}{\partial \theta_k}, \qquad j=1,\dots,p. $$
From the following equations, 
\begin{align*}
\dot{Z}_{kj} & =\frac{\partial}{\partial t}(Z_{kj})=\frac{\partial}{\partial t}\left(\frac{\partial x_j}{\partial \theta_k}\right)=\frac{\partial}{\partial \theta_k}\left(\frac{\partial x_j}{\partial t}\right) \\
& = \frac{\partial}{\partial \theta_k} f_j(\bfx(t;\btheta,\bfx_0),t,\btheta) \\
& =  \frac{\partial f_j}{\partial \theta_k} + \sum_{\ell=1}^{p}\frac{\partial f_j}{\partial x_\ell} \frac{\partial x_\ell}{\partial \theta_k},
\end{align*}
we get another ODE system for the Jacobian of $\bfx(t; \btheta,\bfx_0)$,
$$ \dot{Z}_{kj}  =  \frac{\partial f_j}{\partial \theta_k} + \sum_{\ell=1}^{p}\frac{\partial f_j}{\partial x_\ell} Z_{k\ell}, \qquad j=1,\dots,p. $$
The initial conditions for the system are 
$$Z_{kj}(0)=0 \quad \text{ for all } j=1,\dots,p,$$ 
from
$$ Z_{kj}(0) =\lim_{\varDelta\theta_k\to0}\frac{x_j(0; \theta_k + \varDelta\theta_k)-x_j(0;\theta_k)}{\varDelta\theta_k}, $$
since $ x_j(0; \theta_k + \varDelta\theta_k)-x_j(0;\theta_k)=0. $

The partial derivative with respect to $\bfx_0=(x_{01},\dots, x_{0p})$ can be obtained in the same way, with the initial condition of $ Z_{\ell j}\equiv\frac{\partial x_j}{\partial x_{0\ell}} $ as follows:
$$ \left\{
\begin{array}{ll}
\text{when }j=\ell,  & x_\ell(0;x_{0\ell} + \varDelta x_{0\ell})-x_\ell(0; x_{0\ell})=\varDelta x_{0\ell}, \text{ so } Z_{\ell\ell}(0)=1, \\
\text{when }j\neq \ell,  & x_j(0; x_{0\ell} + \varDelta x_{0\ell})-x_j(0; x_{0\ell})=0, \text{ so } Z_{\ell j}(0)=0.
\end{array}\right. $$

\subsection{Hessian of $\bfx(t; \btheta,\bfx_0)$ with respect to $\btheta$ and $\bfx_0$}
In the same way as above, we can obtain an ODE system about the Hessian of $\bfx(t; \btheta,\bfx_0)$.

\bigskip
\noindent When we define
$$  W_{rk}^{j}\equiv\frac{\partial Z_{kj}(t)}{\partial \theta_r} = \frac{\partial^2 x_j(t)}{\partial \theta_r\partial\theta_k}, $$

\begin{align*}
\dot{W}_{rk}^{j} & =\frac{\partial}{\partial t}\left(\frac{\partial Z_{kj}(t)}{\partial \theta_r}\right)=\frac{\partial}{\partial \theta_r}\left(\frac{\partial Z_{kj}(t)}{\partial t}\right) \\
& = \frac{\partial}{\partial \theta_r} \left[ \frac{\partial f_j}{\partial \theta_k} + \sum_{\ell=1}^{p}\frac{\partial f_j}{\partial x_\ell} Z_{k\ell} \right] \\
&\text{ using that }\big[\quad\big] \text{ is a function of }(\bfx(t; \btheta,\bfx_0),t,\btheta,\bfZ(t; \btheta,\bfx_0)),\\
& = \left[ \frac{\partial^2 f_j}{\partial\theta_r\partial \theta_k} + \sum_{\ell=1}^{p}\frac{\partial^2 f_j}{\partial\theta_r\partial x_\ell} Z_{k\ell}\right] + \left[\sum_{s=1}^{p} \frac{\partial^2 f_j}{\partial x_s\partial \theta_k}  \frac{\partial x_s}{\partial \theta_r} + \sum_{s=1}^{p} \sum_{\ell=1}^{p}\frac{\partial^2f_j}{\partial x_s\partial x_\ell} \frac{\partial x_s}{\partial \theta_r} Z_{k\ell} \right] + \left[\sum_{\ell=1}^{p}\frac{\partial f_j}{\partial x_\ell} \frac{\partial Z_{k\ell}}{\partial \theta_r}\right] \\
& = \left[ \frac{\partial^2 f_j}{\partial\theta_r\partial \theta_k} + \sum_{\ell=1}^{p}\frac{\partial^2 f_j}{\partial\theta_r\partial x_\ell} Z_{k\ell}\right] + \left[\sum_{s=1}^{p} \frac{\partial^2 f_j}{\partial x_s\partial \theta_k}  Z_{rs} + \sum_{s=1}^{p} \sum_{\ell=1}^{p}\frac{\partial^2f_j}{\partial x_s\partial x_\ell} Z_{rs} Z_{k\ell} \right] + \left[\sum_{\ell=1}^{p}\frac{\partial f_j}{\partial x_\ell} W^\ell_{rk}\right].
\end{align*}

\bigskip
\noindent When we define
$$  W_{rk}^{j}\equiv\frac{\partial Z_{kj}(t)}{\partial x_{0r}} = \frac{\partial^2 x_j(t)}{\partial x_{0r}\partial\theta_k},  $$

\begin{align*}
\dot{W}_{rk}^{j} & =\frac{\partial}{\partial t}\left(\frac{\partial Z_{kj}(t)}{\partial x_{0r}}\right)=\frac{\partial}{\partial x_{0r}}\left(\frac{\partial Z_{kj}(t)}{\partial t}\right) \\
& = \frac{\partial}{\partial x_{0r}} \left[ \frac{\partial f_j}{\partial \theta_k} + \sum_{\ell=1}^{p}\frac{\partial f_j}{\partial x_\ell} Z_{k\ell} \right] \\
&\text{ using that }\big[\quad\big] \text{ is a function of }(\bfx(t; \btheta,\bfx_0),t,\btheta,\bfZ(t; \btheta,\bfx_0)),\\
& =  \left[\sum_{s=1}^{p} \frac{\partial^2 f_j}{\partial x_s\partial \theta_k}  \frac{\partial x_s}{\partial x_{0r}} + \sum_{s=1}^{p} \sum_{\ell=1}^{p}\frac{\partial^2f_j}{\partial x_s\partial x_\ell} \frac{\partial x_s}{\partial x_{0r}} Z_{k\ell} \right] + \left[ \sum_{\ell=1}^{p}\frac{\partial f_j}{\partial x_\ell} \frac{\partial Z_{k\ell}}{\partial x_{0r}}\right] \\
& =  \left[\sum_{s=1}^{p} \frac{\partial^2 f_j}{\partial x_s\partial \theta_k}  Z_{0rs} + \sum_{s=1}^{p} \sum_{\ell=1}^{p}\frac{\partial^2f_j}{\partial x_s\partial x_\ell} Z_{0rs} Z_{k\ell} \right]+ \left[ \sum_{\ell=1}^{p}\frac{\partial f_j}{\partial x_\ell} W^\ell_{rk}\right].
\end{align*}

\bigskip
\noindent When we define
$$  W_{rk}^{j}\equiv\frac{\partial Z_{0kj}(t)}{\partial x_{0r}} = \frac{\partial^2 x_j(t)}{\partial x_{0r}\partial x_{0k}},  $$

\begin{align*}
\dot{W}_{rk}^{j} & =\frac{\partial}{\partial t}\left(\frac{\partial Z_{0kj}(t)}{\partial x_{0r}}\right)=\frac{\partial}{\partial x_{0r}}\left(\frac{\partial Z_{0kj}(t)}{\partial t}\right) \\
& = \frac{\partial}{\partial x_{0r}} \left[ \sum_{\ell=1}^{p}\frac{\partial f_j}{\partial x_\ell} Z_{0k\ell} \right] \\
&\text{ using that }\big[\quad\big] \text{ is a function of }(\bfx(t; \btheta,\bfx_0),t,\btheta,\bfZ(t; \btheta,\bfx_0)),\\
& = \left[ \sum_{s=1}^{p} \sum_{\ell=1}^{p}\frac{\partial^2f_j}{\partial x_s\partial x_\ell} \frac{\partial x_s}{\partial x_{0r}} Z_{0k\ell} \right] + \left[\sum_{\ell=1}^{p}\frac{\partial f_j}{\partial x_\ell}\frac{\partial Z_{0k\ell}}{\partial x_{0r}}\right] \\
& = \left[ \sum_{s=1}^{p} \sum_{\ell=1}^{p}\frac{\partial^2f_j}{\partial x_s\partial x_\ell} Z_{0rs} Z_{0k\ell}\right]+ \left[\sum_{\ell=1}^{p}\frac{\partial f_j}{\partial x_\ell} W^\ell_{rk}\right] .
\end{align*}

\bibliographystyle{dcu}
\bibliography{laplace}

\end{document}